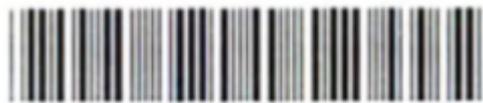

# 硕士学位论文

## 英国配电网长期增量成本定价模型分析

## 及对中国增量配电网启示

**Long Run Incremental Cost (LRIC) Distribution Network Pricing in UK, advising China's Distribution Network**

**Asad Mujeeb**

**2020 年 03 月**

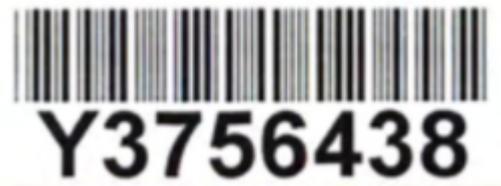

国内图书分类号：TM715 　　　　　　　　　学校代码：10079

国际图书分类号：621.3 　　　　　　　　　　密级：公开

硕士学位论文

# 变低速风电机组输出特性仿真平台的设计

硕 士 研 究 生 ：ASAD MUJEEB 阿萨德

导 师 ：王鹏

申 请 学 位 ：工学硕士

学 科 ：电气工程

专 业 ：电力系统及其自动化

所 在 学 院 ：电气与电字工程学院

答 辩 日 期 ：2020 年 03 月

授 予 学 位 位 ：华北电力大学

I

Classified Index: TM715

U.D.C: 621.3

Thesis for the Professional Master Degree

# Long Run Incremental Cost (LRIC) Distribution Network Pricing in UK, advising China's Distribution Network

| | |
|---|---|
| **Candidate：** | Asad Mujeeb |
| **Supervisor：** | Prof. Wang Peng |
| **Academic Degree Applied for：** | Master of Engineering |
| **Specialty：** | Power System and Automation |
| **School：** | School of Electrical and Electronic Engineering |
| **Date of Defense：** | March, 2020 |
| **Degree-Conferring-Institution：** | North China Electric Power University |

# 华北电力大学硕士学位论文原创性声明

本人郑重声明:此处所提交的硕士学位论文《**英国配电网长期增量成本定价模型分析及对中国增量配电网启示**》,是本人在导师指导下,在华北电力大学攻读硕士学位期间独立进行研究工作所取得的成果。据本人所知,论文中除已注明部分外不包含他人已发表或撰写过的研究成果。对本文的研究工作做出重要贡献的个人和集体已在文中以明确方式注明。本声明的法律结果将完全由本人承担。

作者签名: 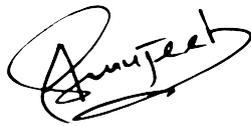　　　　　　　　　　日期: 2020 年 05 月 27 日

# 华北电力大学硕士学位论文使用授权书

《**本人郑重声明:此处所提交的硕士学位论文**》系本人在华北电力大学攻读所示学位期间在导师指导下完成的硕士学位论文。本论文的研究成果归华北电力大学所有,本论文的研究内容不得以其它单位的名义发表。本人完全了解华北电力大学关于保存、使用学位论文的规定,同意学校保留并向有关部门送交论文的复印件和电子版本,允许论文被查阅和借阅。本人授权华北电力大学,可以采用影印、缩印或其他复制手段保存论文,可以公布论文的全部或部分内容。

本学位论文属于(请在以上相应方框内打"√"):

保密口,在年解密后适用本授权书

不保密:✓

作者签名: 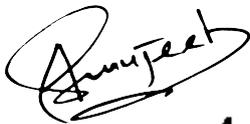　　　　　　日期: 2020 年 05 月 27 日

导师签名: 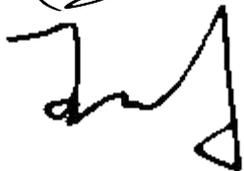　　　　　　日期: 2020 年 05 月 27 日

# 声明

本人郑重声明:此处所提交的硕士学位论文是在华北电力大学攻读硕士学位期间独立进行研究工作所取得的成果,不包含他人已发表或撰写过的研究成果。

作者签名: 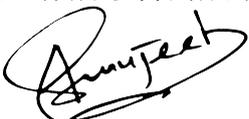　　　　　　　　日期: 2020 年 05 月 27 日

证明人:

导师签名: 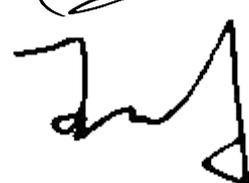　　　　　　　　日期: 2020 年 05 月 27 日

## Declaration and Certification

I declare that this thesis work has been carried out by myself and is unaided work, and is not copied from or written in collaboration with any other person.

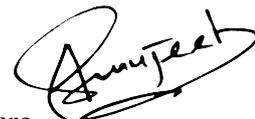

Signature……………………　　　　　　　　Date…2020-05-27….

**Certification By:**

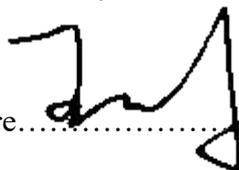

Signature………………………　　　　　　　Date…2020-05-27….

**Acknowledgment**

North China Electric Power University

Department of Electrical and Electronic Engineering

The undersigned hereby certify that they have read and recommend to the Faculty of

Electrical Engineering, for acceptance a thesis

entitled

**Long Run Incremental Cost (LRIC) Distribution Network Pricing in UK, advising China's Distribution Network**

by

**Asad Mujeeb**

in partial fulfillment of the requirements for the degree of

**Master of Science Electrical Engineering**

Dated: March, 2020

**Supervisor: Wang Peng**

**MSc Thesis Committee**

**Prof.___________________**                    **Prof.___________________**

**Prof.___________________**                    **Prof.___________________**

# 摘要

配电网系统是现代电力系统的重要组成部分。近年来，随着能源需求不断增加，可再生能源在电力系统中的渗透率不断提高。越来越多的分布式发电机组（DGs, Distributed Generators）接入配电网，以满足电力系统用户的用电需要。

目前，大量分布式发电机组的接入促使配电网系统逐渐向现代化、智能化演进，也推动了分散式电力市场的发展。英国政府向国内 14 家配电网络运营商（DNO）施压，要求其在配电网系统中接入更多的分布式发电机组，并明确提出指标，要求到 2020 年时可再生能源机组发电量占全国发电量的 15%，到 2050 年提高到 80%。分布式发电机组在配电网中大量应用可以使电网更加绿色、提高电网供电可靠性，但从长远来看，可再生能源的波动性和间歇性也为配电网安全运行带来了诸多挑战。配电网定价对于电网建设非常重要，它保证了电网增量投资后的收益和成本回收。其中，电网运行安全是影响配电网定价计算和分配的重点考虑因素。若电力系统中造成了 N-1 的突发情况，消费者和供应商无法获知网络运行情况，配电网定价问题就变得非常棘手。

为了解决这一问题，本文研究在英国配电网长期增量成本（LRIC, long run incremental cost）定价模型的基础上对网络安全进行了分析。本研究提出一种新的深度强化学习方法，称为深度强化学习算法（DQN, deep reinforcement learning algorithm），以优化网络中的无功功率值，在保持网络安全的前提下，平衡并降低网络定价价格。该方法以 IEEE 14 节点系统为数学模型，用 DQN 算法伪码在 MATLAB 中进行了实际仿真。分析了节点注入网络前后有无安全因素的网络安全。通过多次迭代优化网络无功功率，最终得到了理想的结果。本文还介绍了中国增量配电改革情况和中国配电定价改革情况，分析了英国配电网定价方法对中国的经验启示。

**关键词**：配电网定价，长期增量成本（LRIC），网络安全，无功功率，深度强化学习，节点注入。

# **Abstract**

Electricity distribution network system is considered one of the key component of the modern electrical power system. Due to increase in the energy demand, penetration of renewable energy resources into the power system has been extensively increasing in recent years. More and more distributed generations (DGs) are joining the distribution network to create balance in the power system and meet the supply and demand of consumers.

Today, large amount of DGs inclusion in the distribution network system has completely modernized power system resulting in a decentralize electricity market. Hence, Government of UK is pressurizing 14 distribution network operators (DNOs) to include more DGs into their distribution network system and sets a country level target of generating 15% energy from renewable resources until 2020 and enhancing to 80% by 2050. DGs inclusion in the network system might be helpful due to many factors, but it creates many challenges for distribution network system in the long term. The network security is realized to be one of the challenge that impact the efficiency of accurate calculation and distribution of network pricing among consumers. Network pricing is very essential for the use of the network in a way that it reflects benefits and cost after the injection of new generation into the network. Network security creates N-1 contingency situation in the network which confuse the network to allocate accurate pricing for consumers and suppliers.

To address the aforementioned issue, this research analysed the network security on the basis of Long run incremental cost (LRIC) pricing to balance and reduce the network pricing for the DNOs in UK. However, this study presented an approach of Deep reinforcement learning (DRL) also called deep reinforcement learning algorithm (DQN) to optimize the reactive power values in the network to balance and reduce the network pricing while keeping the network security. The method considers IEEE14 bus as its mathematical model and practically simulates the method in MATLAB using DQN algorithm pseudo codes. The network security has been analysed with and without security factor before and after the nodal injection into the network. Network's reactive power has been optimized through several iterations to get the desired results. Moreover, the research work pointed out the incremental distribution network reforms and its quick implementation in China; lesson learned from the electricity distribution network of Great Britain.

# Contents

# Chapter 1 Introduction

The demand for electrical energy is increasing day by day because of its abundant usage. Recently, the enormous increase in the energy demand along with the deregulation of the power industry has put a lot of pressure on power system operators. The cost and economic impact of blackouts are also increasing and it has become mandatory for utilities to supply safe and reliable power to all the customers. This leads to build a competitive electricity market by transformation of energy sector from a monopoly environment [1].

In Great Britain, the transmission and distribution network system was significantly prolonged in the course of late 1950s to 1960s. Electricity market structure became more mature due to the deregulation of energy industry not only in UK but in the whole world [2]. Reducing cost and enhancing the efficiency has always been proved a major challenge for the electricity power industry. Considering that challenge, the UK electricity industry adopted privatization in 1990; that purely means to enhance the quality of power supply, reliability, effectiveness and meeting consumers demand and supply promoting a decentralized energy environment [3].

In Great Britain, electricity is distributed by the distribution network operators while carrying generated electricity from high voltage transmission lines to the low voltage commercial, industrial and domestic sector. There are total 14 distribution network operators (DNOs). The fourteen distribution network operators are regulated and under licensed by the Office of Gas and Electricity Market (Ofgem) [4]. Ofgem is a non-ministerial organization that functions to regulate the monopoly companies responsible for the transmission, distribution and gas network in UK.

It is responsible for the interests of consumers to provide efficient pricing by keeping tabs on the distribution network operators of gas and electricity distribution along with managing supply, demand and generation. The 14 DNOs are further owned by six different operators divided into different regions of whole Britain [3-5]. Each DNOs covers separate geographical region in UK. DNOs are supported by Independent distribution network operators (IDNOs) that are located in the region allocated to respective DNO. These are small network operators regulated by natural monopolies (Ofgem considered all 14 DNOs as natural monopolies). The structure of 06 operators owned by 14 DNOs are shown in the following figure 1-1 [4-5].

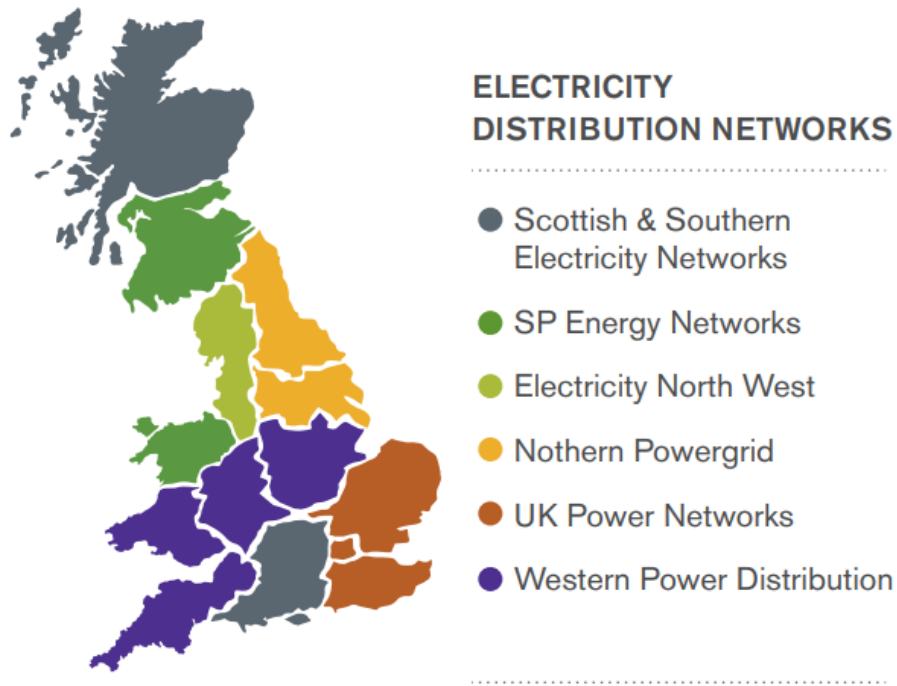

Fig. 1- 1 The 14 DNOs owned by Six Operators in UK

The distribution network operators supply electricity to their customers in their respective region. DNOs in return charge the customers for using their network. Ofgem encourage DNOs operators to charge the customers generating revenues on the basis of connection charges. These DNOs charge the customers using distribution use of system (DUoS) charging methodology which is a prominent and most common methodology for all the 14 distribution network operators in whole UK.

Generators and suppliers are responsible for the payment of DUoS charges for efficient network reinforcement and maintenance to recover the maintenance cost. On the other hand, consumers and generators cover the connection charges for promoting new connections in the distribution system [6]. Ofgem narrated that electricity network are for their customers and DNOs are the natural monopolies functioning by customers to guard every consumer in the network from their dominant control [6-7]. This is described in the figure 1-2.

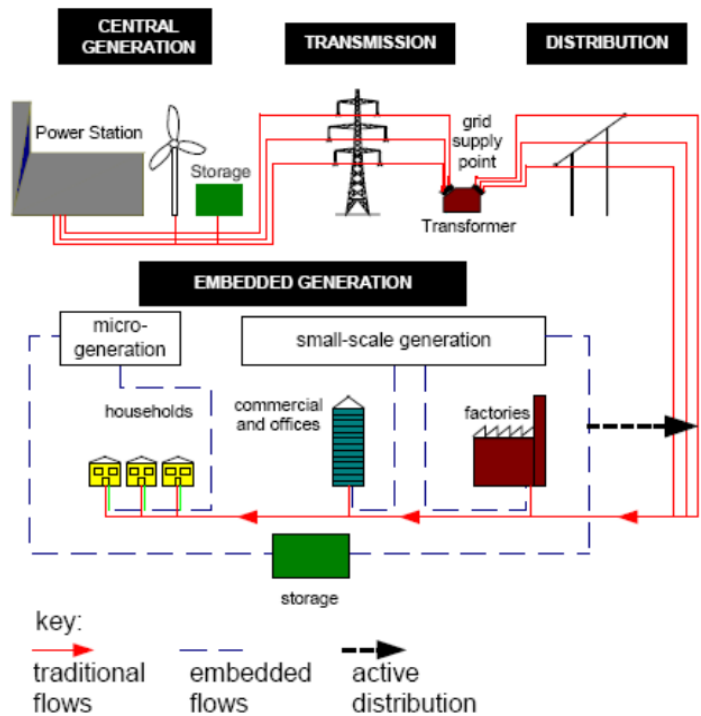

Fig. 1- 2 An Electrical Network for their Customer

The increase in electricity demand and rapid changes in the electricity industry has created lot of challenges for the distribution network operators to minimize electricity cost for the current and future consumers and reducing carbon footprints. Some of the changes in the modern electric industry are due to the introduction of distributed generations (DGs) to tackle the supply and demand issues in the decentralized system. The modern distribution network is explained in figure 1-3 [8].

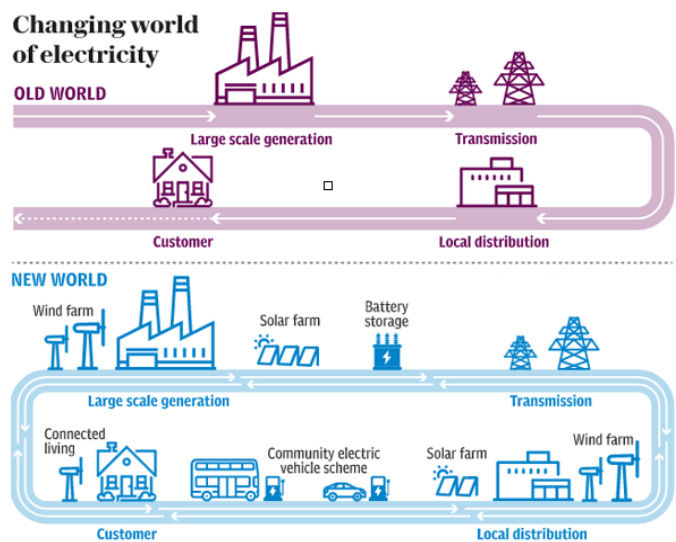

Fig. 1- 3 Modern electricity system with inclusion of DGs in UK

DGs inclusion has modernized the UK electricity power system to generate clean energy from renewable energy resources creating a clean energy environment [7-9]. The government of UK is pressurizing to install more DGs into the distribution network system so that to stabilize the generation matching the demand and supply of consumers. Hence, the Government of UK favoured the increase in production of more renewable energy and sets a country level target of generating 15% energy from renewable resources until 2020. Accompanied by this, the UK government wants more consistency from the DNOs and IDNOs to promote the renewable energy resources so that to fully merge and produce 80% of electricity from renewable and clean energy resources until 2050 [9].

## 1.1    Distribution Network pricing in UK

The advantages of injecting distributed generations in modern power market over the traditional system in the distribution network helps to lower the capital costs, lowering the network charges, reduce carbon footprints because of the clean energy production. This new revolution of DGs need to be adjusted in the current electricity network that creates several challenges, just like technical challenges, regulatory and commercial challenges to better supply reliable power to the consumers. This study defines such type of challenge as the network reinforcement challenge facing by all the 14 DNOs in the UK power sector [10]. In this course, with the passage of time different models have been proposed and applied in the British distribution sector. Distribution reinforcement model (DRM) was the first introduced early in 1984 for calculating distribution network charges. This model was termed as an allocation model because the prices remain fixed and didn't change with location [11]. It attributes the costs of the existing networks to users depending upon the use of each voltage level of the distribution system, as inferred from their maximum demand and customer class characteristics.

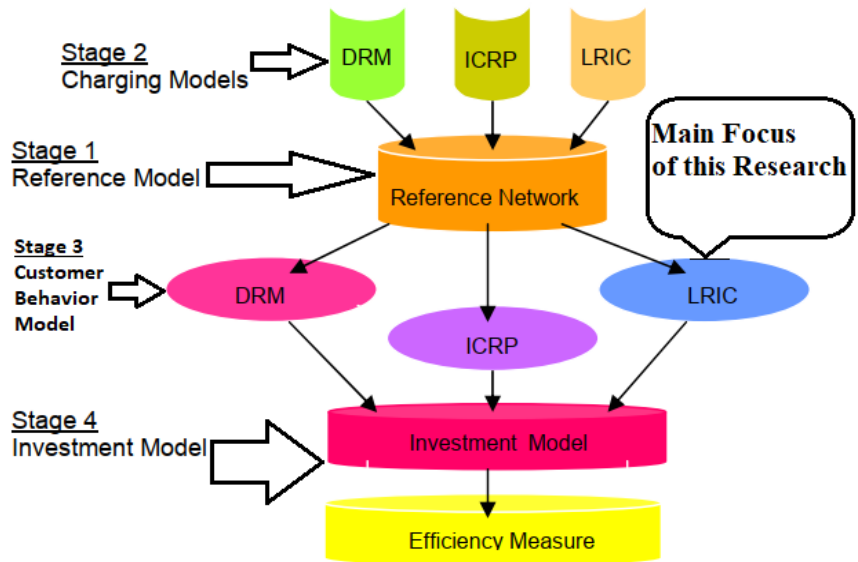

Fig. 1- 4 Pricing Models in UK over the years

Due to the lack of locational signals, DRM model proved to have drawbacks of efficiently promoting competition failing to make a decentralized competitive environment. This greatly affects the customer pricing concerns, lack of efficient power distribution and unable to support the potential increase in the embedded generation [12]. Investment Cost Related Pricing (ICRP) implemented in 2004 as a successor of DRM to fix the locational prices. In this model prices are based on nodal generation and change with voltage and location. Long Run Incremental Cost pricing (LRIC) the most advance method used for the future cost pricing is the latest technique used in UK distribution network system under Ofgem [13].

## 1.2    Long Run Incremental Cost Network Pricing (LRIC)

As stated above, long run incremental distribution network pricing methodology is the current most advanced nodal generation based pricing methodology that can vary both with the location and voltage. It is the best method to reflect the future network pricing affecting upon the consumers and distribution network companies. The LRIC model naturally produces both generation and demand cost irrespective of their symmetry and the charges reflect the need of future investment to meet the incremental generation, demand and supply. This method is the best approach dealing and prediction the future incremental cost and generation and much helpful for embedded DGs generation into the distribution network. All the distribution network pricing are the reinforcement models including LRIC but with different characteristics. LRIC a reinforcement model responsible for long run investment cost is the first model highlighted by this research that links the generation or demand on nodal basis to change in circuits and transformers investment horizon [14-15].

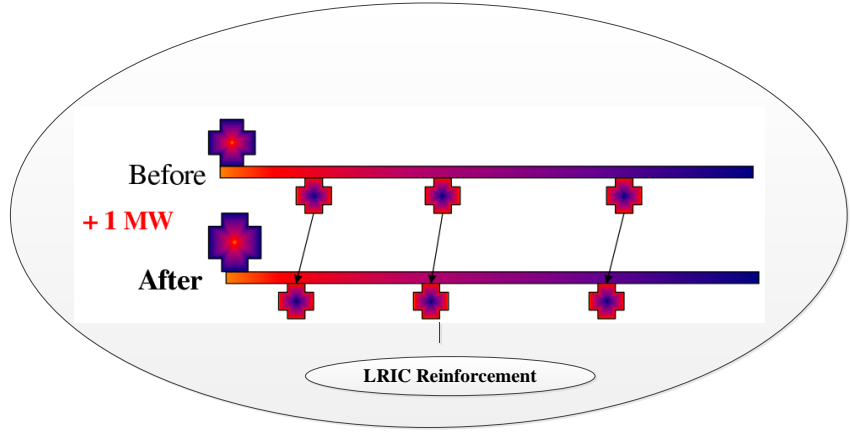

Fig. 1- 5 LRIC reinforcement with load and generation increment

A 1MW load is injected as shown in the figure to produce the reinforcement cost for future investment. This study depicted that LRIC is focused on the asset cost to match the incremental demand and generation and degree of utilisation. Time to reinforce or reinforcement time horizon has a major role in LRIC pricing model and it is denoted by $n_l$. The time to reinforce is the time taken for the load for a given load growth $r_l$ to develop from existing loading level $D_l$ to a specified loading level $C_l$. It is shown in the equation

$$C_l = D_l * (1 + r_l)^{n_l} \qquad (1-1)$$

Taking log on both side and rearranging the equation, we get the time horizon or the time to reinforce

$$n_l = \frac{log C_l - log D_l}{log (1 + r_l)} \qquad (1-2)$$

The circuit long run incremental cost with and without the incremental load is represented in the shape of variation occurs in the present value which is denoted by $PV_l$ which is formulated as;

$$PV_l = \frac{AssetCost_l}{(1+d)^{n_l}} \qquad (1-3)$$

An injection of 1MW at node $N$, can cause to raise the current power flow value to $\Delta P_l$ which will results in a plus new time to reinforcement values and it can be represented as $n_{l\,newly\,added}$ for further reinforcement. This change in the present value is represented in the below equation (1-4). $AssetCost_l$ is the cost of asset investment in equation (1-3). The value "d" in equation (1-4) actually represents the discounted value in a condition where the power flow gets reversed from the future investment condition to the present value $PV_l$ [16].

$$\Delta PV_l = PV_{l \, newly \, added} - PV_l \qquad (1\text{-}4)$$

The change in the present value is formulated in this equation (1-4). So further simplification with respect to discounted investment can leads to equation;

$$\Delta PV_l = Asset \, Cost_l \left( \frac{1}{(1+d)^{n_l \, newly \, added}} - \frac{1}{(1+d)^{n_l}} \right) \qquad (1\text{-}5)$$

Hence, from these equations LRIC can be formulated for the N node injection to the summation of changes made in the injected circuit for current value of investment $\sum \Delta PV$ divided by the nodal injection $\Delta P_{iN}$ multiplied by the total economic lifetime of an asset represented as an annuity factor. The final formula for LRIC pricing calculation is derived here in equation (1-6)

$$LRIC_N = \frac{\sum \Delta PV}{\Delta P_{iN}} * Annuity \, factor \qquad (1\text{-}6)$$

## 1.3 Research Problem

As explained, the LRIC pricing model is the advance methodology supporting distribution network pricing and overcomes the shortcomings of both DRM and ICRP, but its implication in the network is still a difficult task because of its complexity. However, it provides an economical signal to the current and future consumers fulfilling the network requirements somehow it is not considered as a perfect method. This study analysed some demerit and the challenges LRIC is facing that make it complex to deploy in practical circumstances. Some of these issues are;

➢ Network Security Factor- Evaluation of maximum marginal capacity before the reinforcement requirement of the network).

➢ Time to Reinforce Evaluation- Accordingly, Circuit loading growth rate.

➢ Revenue Reconciliation- Evaluate and generate the LRIC final tariff.

This research is based on the "Network security factor" faced by the LRIC pricing technique in UK. This study deeply analysed the network security factor as one of the important aspect in calculating distribution network pricing. Integration of renewable generation resources into the network might be beneficial for fulfilling the generation and demand for the consumers but it creates lot of challenges in the shape of imbalance and system security for the distribution network and distribution network operators. In doing so, this thesis research suggested Deep reinforcement model (DRL) to efficiently and effectively calculate the long run incremental cost pricing for UK distribution network.

On the other hand, China's distribution network is facing massive price distribution challenges. This suggested DRL methodology is also a base model for China's distribution network pricing market to efficiently implement such type of pricing technique in the constructed Pilot projects implemented by China's State grid and China's Southern power grid around the country to challenge the monopoly system charges. This will make the functioning of Pilot projects much more beneficial in reducing network distribution electricity cost and will motivate building more Pilot projects in future.

## 1.4    Contribution of this Study

The key contributions of this thesis are gives as:

- ➢ A detail study of distribution network structure in Great Britain.
- ➢ Summarize overview of different types of distribution network pricing used in UK over the years and its mutual comparison based on different factors.
- ➢ Developing deep understanding of the most advanced Long Run Incremental Cost (LRIC) Pricing highlighting its benefits and challenges in UK network system.
- ➢ In depth study of China's Electrical power distribution network system, its contribution and advancement and role of Pilot projects in improving and tackling the distribution network pricing.

In doing so, this thesis attempts to:

- ➢ This research introduces Deep reinforcement learning (DRL) technique to tackle the network security problem in distribution network.
- ➢ To determine the mathematical model of reactive power optimization and select the appropriate deep reinforcement learning algorithm (DQN).
- ➢ To optimize and minimize the reactive power flow in the network by using iterative method via DRL mechanism to regulate the network pricing
- ➢ To undertake the power flow analysis that includes the N-1 contingency analysis condition in the distribution network.
- ➢ Adjust the parameter, optimized the reactive power optimization algorithm model and train the model, and test the performance of the trained model.
- ➢ To compare the reactive power calculated after applying Deep reinforcement learning (a deep neural network) alongside each iteration results to identify the decrease/increase in the LRIC pricing with and without security factor (S.F).
- ➢ Demonstrate the results and apply it to China's distribution network charges calculation.

## 1.5 Layout of this Thesis

This research thesis includes six chapters which are arranged in the following manner.

**Chapter 1** of the research describes the basic introduction of electrical power system in Great Britain, Distribution Network pricing and primary focus on Long Run incremental cost (LRIC) distribution network pricing in UK. Alongside, the major contribution of this research also includes pointing out challenges faced by LRIC pricing.

**Chapter 2** is based on the literature of this research that discusses in detail the basic and advance electrical power system and embedded generation such as distributed generations (DGs) in the modern power system, Distribution network structure in UK, LRIC in UK and its basic formulation and Electrical distribution network structure in China. Literature of Distribution reinforcement learning (DRL) and Current pricing methods used in the China's distribution network is also discussed in detail.

**Chapter 3** addresses the research methodology of this study for LRIC pricing. The research methodology proposes the Deep Reinforcement Learning (DRL) technique to optimize the reactive power values while regulating the pricing by continuous iterations. This section of the thesis contributes to present power flow in the distribution network and analysed three approaches dealing with the distribution network system model. This section proposes DRL model and its significance evaluating the reactive power with the help of IEEE-14 bus system while minimizing the LRIC pricing for future consumers.

**Chapter 4** deals with the results analysis and discussion as an output from the whole research. The results are calculated and applied on the distribution network to tackle the network security challenge faced by LRIC pricing in UK. The results analysed the pricing difference in LRIC model on the basis of nodal injection in the network and compared the network security factor with and without nodal addition. The results are discussed in details on the basis of practical values created from the applied methodology.

**Chapter 5** highlights the incremental distribution network reforms in China. This section of the research sheds light on the current distribution network system in china and the reform that has been brought up to the incremental distribution power market until now. The study provides suggestions from the Great Britain experience of Incremental distribution network both on the basis of long term and short term.

**Chapter 6** presents the conclusion of this study and its future perspectives.

# Chapter 2 Literature Review

This Chapter presents basic literature review about Electrical Power System, main concept about distribution network system, Long Run Incremental Cost Pricing, distribution network pricing in UK and their effect on distribution network market mechanism. Deep reinforcement learning network and their basic function is also highlighted. Last but not least, the distribution network implementation and their current pricing methods in China's distribution energy market are also discussed.

## 2.1 Electrical Power System

The structure of a traditional electric power system consists of three main sections: generation, transmission and distribution that are clearly distinguished by either stepping-up or stepping-down power transformers. Figure 2-1 presents the structure of a typical electric power system.

Generation consists of large power stations, which have a typical size of higher than 500 MW power, produced at around 20 kV. The biggest share of the electricity generation is produced by a small number of large fossil fuel fired power stations where high temperature is used to produce high pressure steam in boilers used to drive the turbines which in turn rotate the large generators' shafts [17]. There are a few countries around the world that possess nuclear power stations for generating electricity (the required high pressure steam is produced by nuclear fission instead of burning fossil fuels), while almost all countries possess generating units that are using renewable energy sources (solar, wind, geothermal, biomass, etc.), since these technologies are considered more environmentally friendly than the traditional sources (coal, oil and natural gas) [18].

Transmission consists mainly of an overhead transmission network (the percentage of underground and submarine cables is relative low) that transfers the electric power from generating units to distribution, which eventually supplies the load [19]. The transmission network is also interconnected with neighboring transmission networks for exchanging electric power due to both financial (electric power trading) and practical reasons (maintenance scheduling, sharing of generation reserves and emergencies, i.e., lack of sufficient electric power, assistance in the grid stability, etc.). Stepping-up transformers are used to raise the voltage of the generated power in order to transmission it to long distances. This has a result the reduction of current for a specific amount of transferred electric power, resulting in the reduction of power losses and in the more cost-effective construction of transmission network's infrastructure (size of conductors, type of towers, etc.).

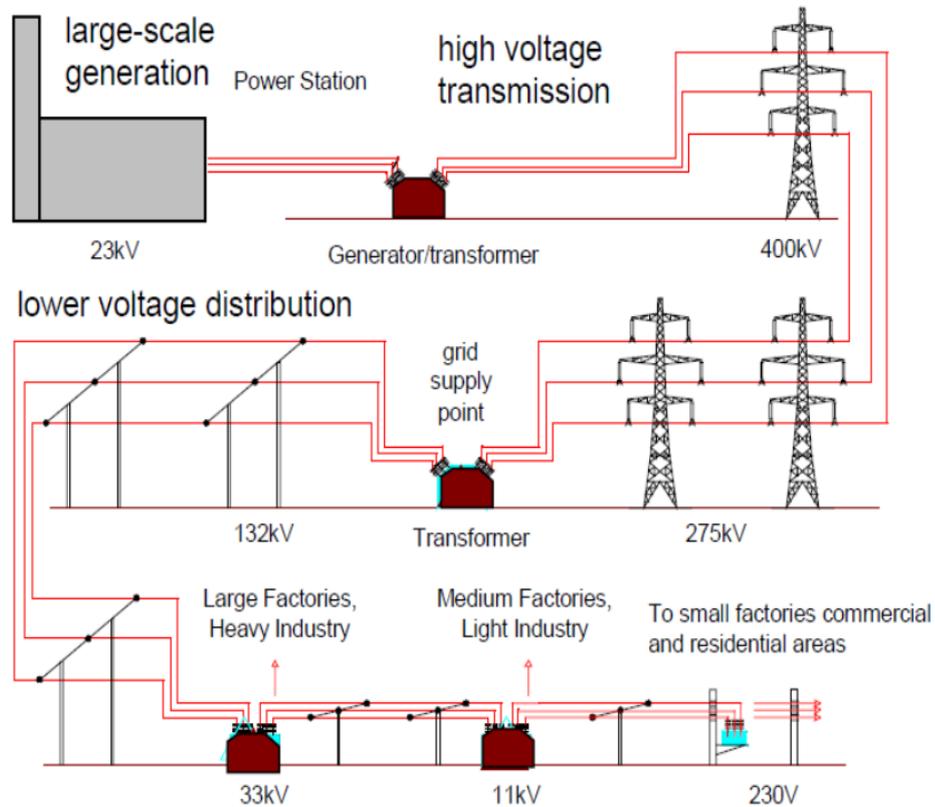

Fig. 2- 1 Basic Electrical Power System [20]

Based on the international standards transmission voltage networks operate at 132 kV or higher voltage. The transmission networks up to 230 kV are usually referred as high voltage (HV) transmission networks, while transmission voltages above 230 kV are usually referred to as extra-high voltage (EHV). Finally voltages above 1000 kV are referred to as ultra-high voltages (UHV). The United Kingdom possesses the following EHV and HV networks: 400 kV and 275 kV in England and Wales, 400 kV, 275 kV and 132 kV in Scotland, and 275 kV and 110 kV in Northern Ireland. Distribution consists of distribution networks and stepping-down transformers that distribute electricity to different consumers. The distribution network links the distribution transformers to the consumers' service-entrance equipment. The primary distribution lines range from 4 to 34.5 kV and supply the load in a well-defined geographical area. Some industrial customers are served directly by the primary feeders. The secondary distribution network reduces the voltage for utilization by commercial and residential consumers. It serves most of the customers at levels of 230 V (Europe)/120 V (North America), single-phase, and 400 V (Europe)/208 V (North America), three-phase. Distribution networks make use of both overhead conductors/cables and underground cables. The growth of underground distribution has been extremely rapid and up to 70% of new residential construction in North America and Europe is via underground systems [21].

With the increasing demand in energy for commercial, industrial, and local sector have created a huge challenge for the electricity market and traditional electrical power system all over the world. Environmental challenges cannot be neglected considering modern electrical power system. These challenges lead a traditional power system into a modern power system introducing distributed generation (DG) and integration of renewable and clean energy into the system [22]. With the passage of time, distributed generation are moving towards advancement in which DGs are divided into the dispatchable and non-dispatchable units [23] as show in the figure below.

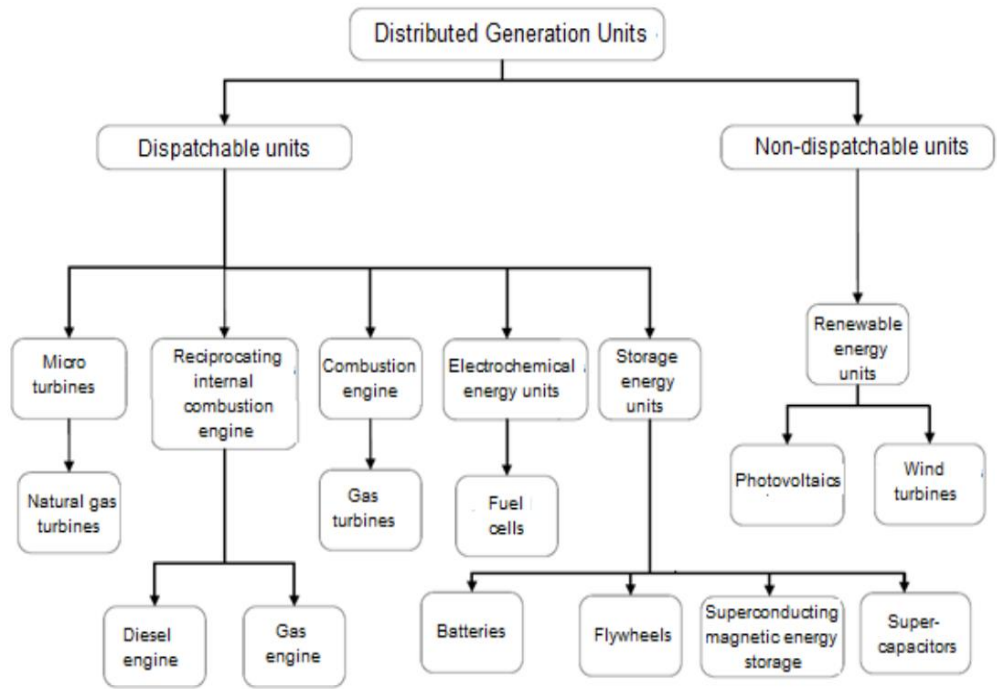

Fig. 2- 2 Distributed generation technologies system in Modern Power System [22-23]

DG units can be used either in an isolated way, supplying power only for serving the consumer's local demand, or in an integrated way, supplying power to the electric power system. In distribution systems, DG offer many advantages for both consumers and electric utilities, especially in cases where central generation is not feasible or in cases where there are serious issues/problems with the transmission network [24]. Figure 2-3 represents an evolution of traditional grid system to the modern power system through DG inclusion.

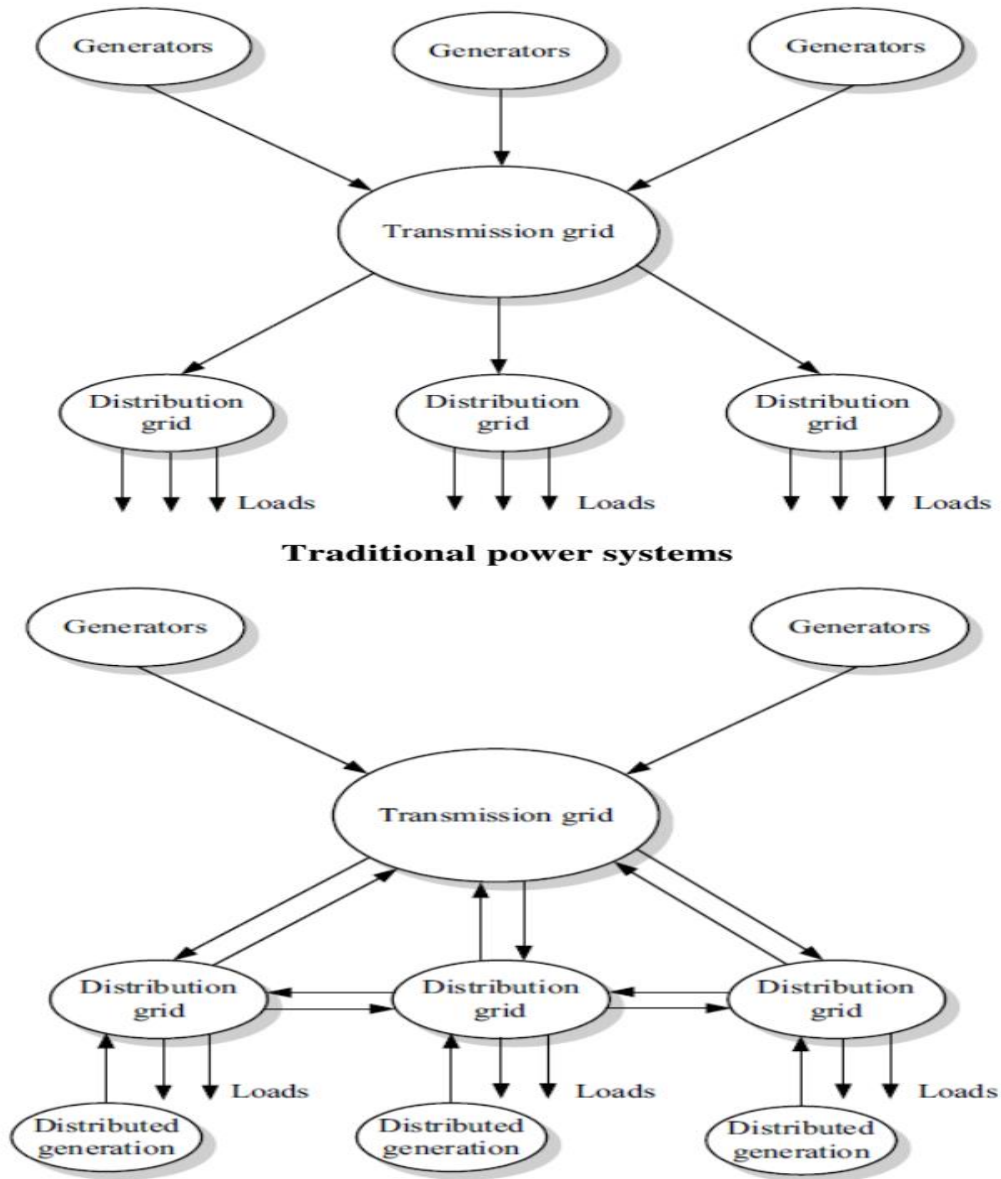

Fig. 2- 3 Traditional VS Modern Power System with DG inclusion

## 2.2 Electrical Distribution Network System (A General Overview)

Electricity is delivered to the customers/consumers through a complex distribution network. Electricity is generated at power plants and moves through a complex system, sometimes called the grid, of electricity substations, transformers, and power lines that connect electricity producers and consumers. Most local grids are interconnected for reliability and commercial purposes, forming larger, more dependable networks that enhance the coordination and planning of electricity supply [25].

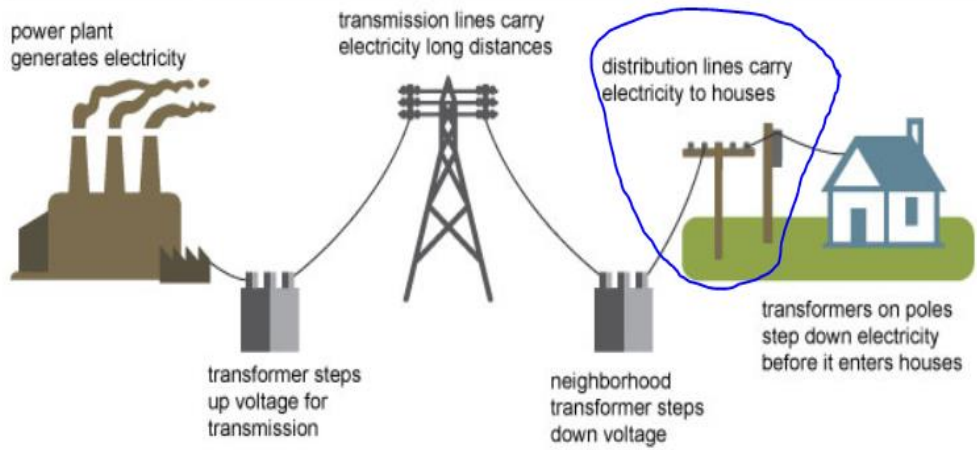

Fig. 2- 4 Electricity structure consisting distribution network

The origin of the electricity that consumers purchase varies. Some electric utilities generate all the electricity they sell using just the power plants they own. Other utilities purchase electricity directly from other utilities, power marketers, and independent power producers or from a wholesale market organized by a regional transmission reliability organization. Power plants generate electricity that is delivered to customers through transmission and distribution power lines. High-voltage transmission lines, such as those that hang between tall metal towers, carry electricity over long distances to meet customer needs. Higher voltage electricity is more efficient and less expensive for long-distance electricity transmission. Lower voltage electricity is safer for use in homes and businesses. Transformers at substations increase (step up) or reduce (step down) voltages to adjust to the different stages of the journey from the power plant on long-distance transmission lines to distribution lines that carry electricity to homes and businesses [25-26].

The employment of renewable energy systems (RESs) such as wind energy systems, photovoltaic (PV) systems, biomass power plants, and small hydro turbines is increasing in electrical distribution networks [27]. Integration of these RESs and other types of distributed generation (DG) units is one of the major changes that may impact load flow, stability, and protection of power systems [28]. Most RESs are interfaced with the grid through power electronic converters. These types of changes are making the distributed network complex and leads to a hybrid network. Hybrid networks with AC grid, AC transmission, DC Grid and combination of DGs, renewable resources are shown in the Figure 2-5 [26-28].

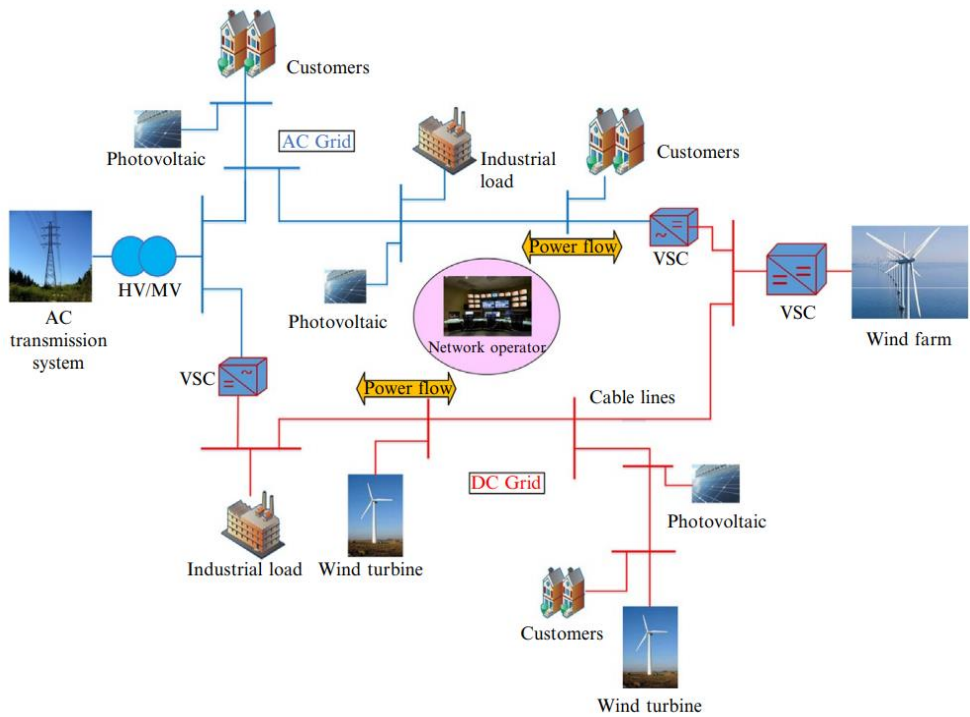

Fig. 2- 5 Future hybrid Distribution Network Architecture [28-29]

## 2.2.1  Distribution Network Structure in UK

The existing electricity networks have been constructed for the predicted patterns of electricity demand and generation. They facilitate the use and generation of electricity in the current particular type of energy system. Great Britain's centralized system of electricity generation involves very large power stations with individually significant environmental impacts. Only a small proportion of national generating capacity is connected to the regional lower-voltage distribution networks. Electricity consumers are passive, rather than responding to changing conditions in the electricity system [30]. Since the privatization of the UK electricity sector, energy policy objectives and regulatory duties have progressively introduced sustainable development issues within an 'unbundled' industry structure. Many countries are facing legislative and commercial pressures to separate or unbundle ownership of electricity networks, supply and generation. The relatively early privatization of Great Britain's electricity supply industry has simply given more experience in this area; other countries face similar challenges [31].

In Great Britain there is a privatized and regulated electricity system, with separate licensed roles for suppliers, for generators, for the bulk transmission networks, for the national balancing system and for the regional distribution networks [32]. After connection, electricity users and generators contract with a supply company as the commercial hub of this system. Companies operate in competition to generate and supply electricity.

Consumers are able to choose their supply company; a distribution network serves all the suppliers operating in that region [33]. Regulation requires supply and distribution companies to operate separately, even where a parent company may own both types of business. The electricity networks operating at 132 kV and below are generally classed as distribution networks [34]. They are owned and operated by 14 regional distribution network operators (DNOs). DNO businesses were formed from the unbundling of the electricity industry from a vertically integrated nationalized system, the result of a process that began with the privatization of Great Britain's electricity industry in 1990 [35]. These private companies are licensed to own and develop the existing network assets for each region [30]. Electricity Network Company's exhibit properties of natural monopoly [36-38]. So the total amount charged by DNOs and transmission companies to network users – through suppliers – is limited by the regulator. The regulatory system is thus very important to DNO businesses.

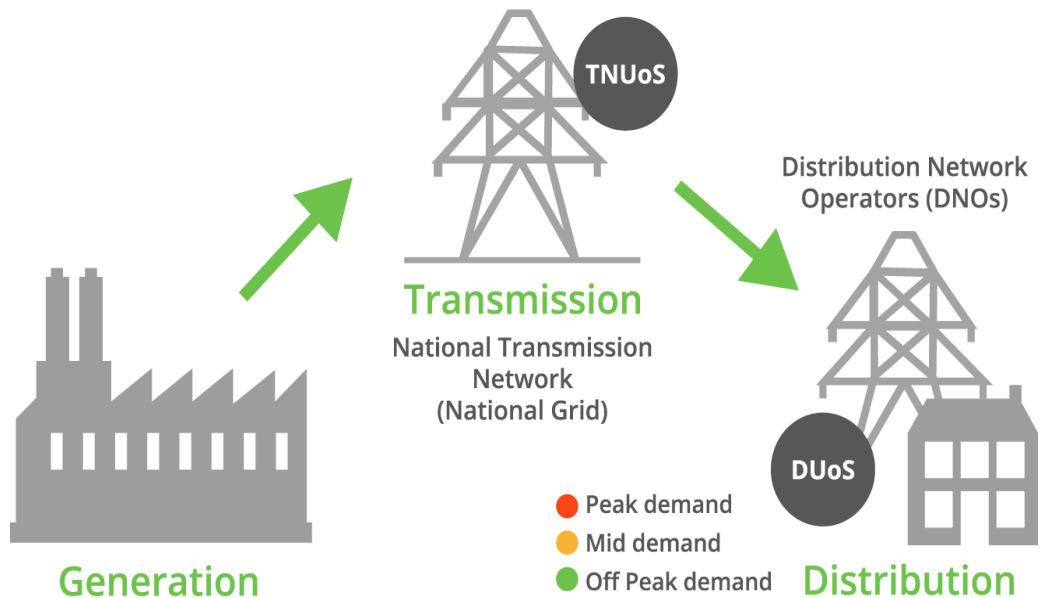

Fig. 2- 6 Electricity structure in UK

Since the Utilities Act 2000, the regulator of the DNOs has been the Office of the Gas and Electricity Markets Authority, Ofgem. Given the system of regulated privately owned distribution networks in Great Britain, society (via the regulator) has an opportunity to set the objectives of the DNO businesses. Thus the role of the regulator – and its interpretation of its role – are important in shaping the role of the networks e.g. the extent to which addressing sustainability is viewed as part of the regulator's role and the networks' function [39]. The framework for treatment of costs and income in the regulatory system is crucial in defining how the DNO business can operate to be profitable. Together with the system of external technical rules and DNOs' own policies, these define the DNO business model.

Although now under review, an RPI-X price-control approach has been applied to Great Britain's electricity networks for 20 years since privatization [40-41]. In this approach, the regulator decides what an efficient DNO needs to spend in total in the next five years to meet its license obligations, and decides on a cost of capital for the income stream from DNOs' investment in regulated assets. As the income is regulated, this reduces the commercial risks to the DNO of recovering their investment, reducing the financing cost. The DNO can invest at a lower cost of capital than otherwise. In an asset-intensive activity such as networks, cost of capital is particularly relevant to overall costs. The regulator can thus restrict the cost of the network service to network users. Ofgem sets each DNO's total allowed income for 5 years, based on its view of operating and investment costs, scaled annually by a consumer measure of inflation (RPI) minus an efficiency factor (X) [42]. This provides strong incentives to reduce costs, since the income is set until the next price control [40-42]. A number of financial incentive schemes vary the basic income allowance of the DNO, based on actual network performance, usage and expenditure [43-44]. These schemes are reset in each periodic review, transferring the benefits of DNOs' efficiency gains to customers through reduced prices, and updating for new information and plans. The features and performance of the RPI-X system have been well explored by many authors, particularly in terms of the difficulty of setting initial allowances and of the experience in Great Britain [45-47]. Use of the RPI-X system in the first three DNO price controls post-privatization (1990–2005) led to a halving of distribution charges, reductions in the frequency and duration of interruptions to supply, increases in investment and reductions in the cost of capital [48]. The regulatory system incentivized the DNOs to reduce costs while maintaining or improving the network service to customers [49-50]. There was innovation in the delivery of services only to the extent that the regulatory system would reward it, so predominantly in relation to reducing costs and supply interruptions; the DNO role was not changing. The regulatory system encouraged DNOs to reduce their workforces, and to restrict maintenance and replacement activities [51]. The distribution regulatory system currently no longer dependent upon the RPI-X scheme. In [42] Ofgem introduced a new distribution network scheme that will run for 08 years of period starting from July, 2015 to April, 2023. Ofgem describe this type of distribution network innovation as "RIIO ED1 electricity distribution network scheme. RIIO stands for Revenue, Incentives, Innovation and Output along with ED1 as Electricity distribution Scheme first. After April 2023, Ofgem will lead the distribution network in UK to the next level of RIIO-ED2 mechanism [42].

## 2.2.2 Long Run Incremental Cost (LRIC) Pricing

Long run incremental cost (LRIC) is a forward-looking cost that a company needs to include in its accounting. Long run incremental costs are gradual costs a company is able to predict and plan for over the long term. Long run incremental cost (LRIC) refers to the changing costs that a company can somewhat foresee. Examples of long-run incremental costs include energy and oil price increases, rent increases, expansion costs and maintenance expenses. Electricity prices offered to customers are generally reflective of generation and network costs involved [52]. Network cost includes transmission and distribution network cost. This cost forms a significant component of electricity prices. Users respond to these network charges by modifying their usage pattern. Efficient network charges invariably reflect the impact of network congestion. Thus, the customers effectively respond to network congestion, caused by supplying power to a set of customers. Transmission pricing models based on nodal pricing mechanism do provide price signals to customers, reflecting the impact of network congestion. However, distribution pricing models differ from transmission pricing models, because of their inherent technical differences, and are unable to provide a true reflection of network congestion [53]. Distribution network prices can be calculated ex-ante through various pricing models viz. Distribution Reinforcement Model, Investment Cost Related Pricing, Long Run Incremental Cost Pricing, and Forward Cost Pricing. LRIC charging mechanism is the most advanced pricing model till date, with a verified potential to save hundreds of millions of pounds in UK, in terms of network investment [54]. This approach is recognized as an economically efficient approach for allocating network cost, as it determines network charges in terms of the difference in present value of future investment, consequent upon nodal power perturbation for generation or demand [55].

Enhanced versions of LRIC methodology consider network security, component reliability and nodal unreliability tolerance [56-57]. This model also respects user's security preferences while assessing their impact on network development cost [58]. In calculating nodal LRIC charges diversity factor is used to calculate the maximum demand at individual locations on network. This factor considers maximum demand of individual users that may not be coincident to network peak demand. Hence, this factor is not able to reflect user's network usage during peak demand [59]. Thus, the approach does not charge users on the basis of their contribution to network peak conditions. Networks are designed to supply peak load on system. Tariff for network services depend on load situation when there is peak demand on system [60]. Each customer category contribution to system peak demand affects network investment, and hence should be reflected in network charges. Network charges are a component of electricity charges. Electricity charges for customers vary with their category classification [61].

**2.2.2.1 Mathematical Formulation**

$$n_c = \frac{logC_c - logD_c}{\log{(1+r)}} \tag{2-1}$$

The current power flow is represented by $D_c$, and capacity $C_c$, load growth with small r.

The future value for the discount rate "d" could be determined as

$$PV_C = \frac{Asset_C}{(1+d)^{n_c}} \tag{2-2}$$

Here, $Asset_c$ denotes asset cost of the network, Load increment has a major effect over cost; hence the nodal injection changes the direction of power flow with related network by $\Delta P_c$ that emerges a new reinforcement horizon below:

$$n_{c\,new} = \frac{logC_c - \log{(D_c + \Delta P_C)}}{\log{(1+r)}} \tag{2-3}$$

In return, it affects the future investment due to the present value

$$PV_{C\,new} = \frac{Asset_C}{(1+d)^{n_c\,new}} \tag{2-4}$$

It result to a new equation due to change in present value

$$\Delta PVc = PVc\,new - PVc \tag{2-5}$$

$$= Asset_C \left(\frac{1}{(1+d)^{n_c\,new}} - \frac{1}{(1+d)^{n_c}}\right) \tag{2-6}$$

For network component C the annualized unit incremental cost is represented as

$$IC_c = \Delta PV_c \times annuity\,factor / C_c \tag{2-7}$$

Finally, the long run incremental cost can be described as

$$LRIC_N = \frac{\sum_c IC_c}{\Delta P_{In}} \tag{2-8}$$

It is the summation of all incremental cost through N node over all the circuits; $\Delta P_{In}$ is the injecting power at the nodes

## 2.2.3 Deep Reinforcement Learning (DRL)

In this thesis, deep reinforcement learning is used for calculation and optimization of reactive power. With the help of iterative approach, reactive power has been optimized several times for different iteration (discussed in details in chapter 3 and 4). After Google's AI research team launched Alpha Go, deep reinforcement learning became a popular algorithm [62-63]. Deep learning achieves approximation of complex functions by training a network with multi-layers of hidden layers, so as to learn the basic characteristics of data sets. Deep reinforcement learning (DRL) is a new algorithm which combines deep learning (DL) derived from neural network with reinforcement learning (RL) in machine learning. This method realizes the learning from perception to decision-making. It can extract and analyze environmental information and guide objects. By inputting data, the deep neural network constructed by deep reinforcement learning can process the data and output actions without manual intervention. Reactive power optimization of power grid is not only a non-linear problem, but also a hybrid problem with "three-plus" characteristics. The so-called "three-plus" refers to multi-variable, multi-objective and multi-constraint. The optimization variables include both discrete variables (number of reactive power compensation devices, transformer gear) and continuous variables (node voltage). These characteristics make optimization a complex problem. At present, the algorithms used in domestic and foreign power grids cannot adapt to the complex grid environment. Real-time automatic optimization control of reactive power and voltage in power grids needs to find a more suitable algorithm to realize automatic optimization of reactive power and voltage. Deep Reinforcement learning is a widely used method, which can adapt to complex environment. It is considered as the most potential method to realize advanced artificial intelligence in the fields of robot control, intelligent driving, process optimization decision-making and control. Deep reinforcement learning has great advantages in dealing with complex and multifaceted problems because it combines perception and decision-making power. Most of the problems in power system are dynamic, multi-constrained and non-linear. Traditional calculation methods have different problems, such as poor convergence or low accuracy. It is flexible and convergent to optimize reactive power in power system by deep reinforcement learning, which can effectively control voltage offset. The deep neural network in deep learning can be classified and predicted, and can be combined with the model-free reinforcement learning algorithm, which can update the control strategy online and realize direct output action without manual intervention. The main algorithms of deep reinforcement learning are shown in table [64].

Table 2- 1 Main Algorithms for Deep Reinforcement Learning

| Algorithm Name | Algorithm Type |
| --- | --- |
| Deep Q Network (DQN) | Based on value function |
| Improved Deep Q Network (Nature DQN) | Based on value function |
| Deep deterministic strategy gradient algorithm (DDPG) | Strategy based |
| Trust Region Strategy Optimization (TRPO) | Strategy based |
| Asynchronous Advantage Actor Critic Algorithms (A3C) | Strategy based |
| Unsupervised Assisted Reinforcement Learning (Unreal) | Strategy based |
| Distributed Approximate Strategy Optimization (DPPO) | Strategy based |

With the increasing speed of computer processors, increasing storage capacity, and diversified forms of data expression, all types of transactions can use deep learning for real-time data analysis.

Deep learning is the use of multilayer perceptron learning models for supervised or unsupervised data learning. The layers in the model are composed of multiple pieces of non-linear data structure transformation. The higher the level, the more abstract the expression of the data features. The learning process of the deep learning model for data is hierarchical. The initialization parameters of the lower layer are derived from the results that have been trained by the upper layer. Then, through a large amount of training, the feature information that should be obtained is accurately obtained to improve the accuracy of learning. Common deep learning models include convolutional neural networks, deep belief networks, and self-coding models. Deep neural network (DNN) adopts the structure of "complex model + simple features" and uses complex non-linear models to learn simple feature relationships. A deep neural network consists of an input layer, an output layer, and multiple hidden layers in between. It is similar to a multi-layer perceptron, except that it has multiple hidden layers, and each hidden layer has multiple interconnected neurons. The advantage of multiple hidden layers of DNN is that it can approach complex decision functions.

The role of the hidden layer is to implement the transformation of complex functions. They are connected to the input layer, label the input values and combine the weights of the input values to produce a brand new, true value, which is then passed to the output layer. The output layer uses the abstract features calculated in the hidden layer for classification or prediction. The advantage of multiple layers is that complex functions can be represented with fewer parameters.

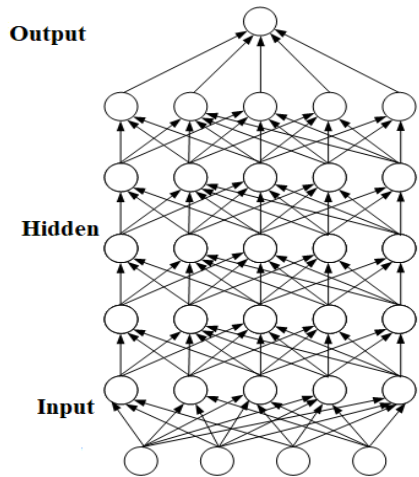

Fig. 2- 7 DNN Network Model

The idea of reinforcement learning has been proposed since the beginning of the 20th century. After nearly a century of development, reinforcement learning has close links with disciplines such as operations research, intelligent control, optimization theory, and cognitive science. It is a typical multidisciplinary interdisciplinary field. The figure below explains the basic principles of reinforcement learning [64-65].

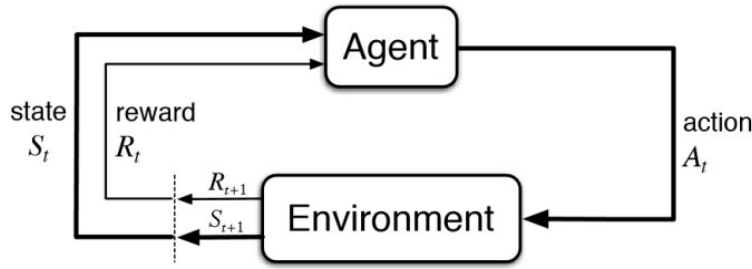

Fig. 2- 8 Basic principles of reinforcement learning

When an agent completes a task, it first interacts with the environment through action "a". The agent will generate a new state under the action of action "a" and the environment. At the same time, the environment will immediately give a return value "r". Reinforcement learning will be based on the new State and reward values and a large amount of data generated during the interaction are calculated and the next action strategy is modified. After tens of thousands of such loop iterations, the trained agent can finally choose an optimal action (optimal strategy) to complete the corresponding task according to each state change.

### 2.2.4  Electrical Distribution Network System in China

The distribution network consists of overhead lines, cables, towers, distribution transformers, disconnections, reactive compensators and some ancillary facilities, supplying and distributing power to users. As the requirement for power supply safety and reliability has been continuously enhanced, urban construction scale and economic development call for high requirements for the distribution network, i.e. relatively perfect multi-loop feeders, good urban planning and electricity path distribution, reliable primary and secondary and network communication equipment.

China's power industry always pays attention by bringing enhancement in the distribution network. The China's National Energy Bureau (NEB) issued a "Distribution Network Construction and Transformation Plan of Action" in 2015 that will run from 2015-2020. This plan focuses on the speeding up the construction of distribution network transformation and intelligent distribution network. It also includes enhancing the adaptability and improving the core competitiveness of the distribution network. However, for long time China lag design standard of distribution network planning work due to the lack of scientific and normative standard system. As the power system industry gradually attach great importance to the construction of distribution network planning, as a result distribution network has released a number of areas for the planning and design standards [66]. At present, China adopts the four level systems, divided into national standards, industry standards, local standards and enterprise standards. In addition, the standard system can also be divided into compulsory standards and recommended standards. Two kinds of standard system design of distribution network in China is relatively mature, the characteristics of the national grid company based design with the construction of distribution network engineering. Figure 2-7 describes China's current distribution network system [67].

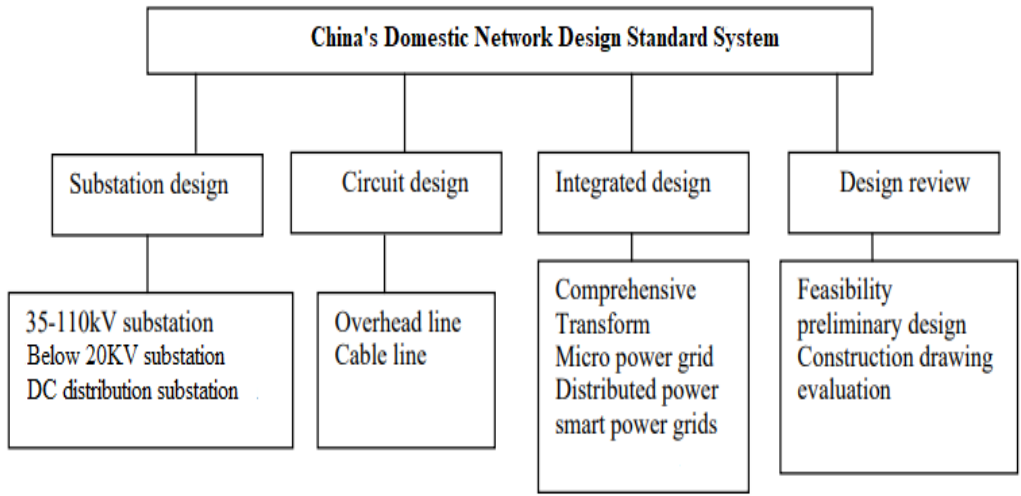

Fig. 2- 9 China's current electricity network distribution design

The distribution network structure determines the reliability and the flexibility of network operation. Various countries have various designs. Taking urban distribution network as an example, the cable network in urban area of Paris is in three-loop network T-connection or double-loop T-connection mode. The cable network in London is in multi-branch and multi-connection mode. In Tokyo, 22kV cable network is built with main lines and backup lines in loop within mesh network, and 6kV overhead network is in multi-section and multi-connection mode, and the cable network is in multi-segmentation and multi-connection mode. The cable network in Singapore is in "petal mode", i.e. looped network in normal closed connection. Despite differences in specific topologies, the advanced grid structure at home and abroad basically tends to be developed in "dumbbell shape", with the core principle of "intensifying the two ends and simplifying the middle", which can guarantee the safety and reliability while avoiding repeated construction [68]. The distribution network in China is classified as high voltage, medium voltage and low voltage according to voltage levels, where high voltage distribution network is often 35-110kV, medium voltage distribution network generally covers 6-10kV and low voltage distribution network is 220/380V. In mega polis with higher load, the 220kV grid can also have distribution function.

It can be classified as urban, rural and factory distribution networks according to the functions of power supply areas [69]. Considering the required level of the power supply security, it is appropriate to have simple and clear structure of the distribution networks, with different grid structures for different power supply areas. In China, the typical cable grid structures mainly include double loop network and single-loop network, as respectively shown in figures below [70].

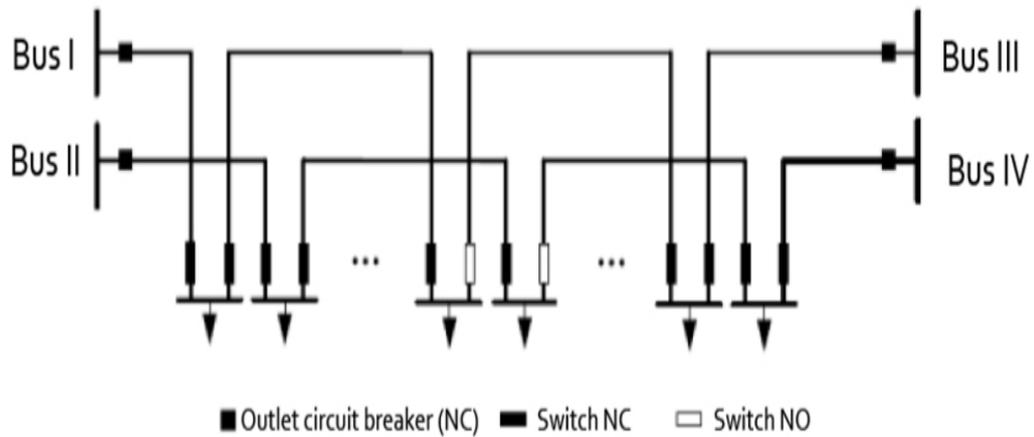

Fig. 2- 10 A 10 KV double loop network

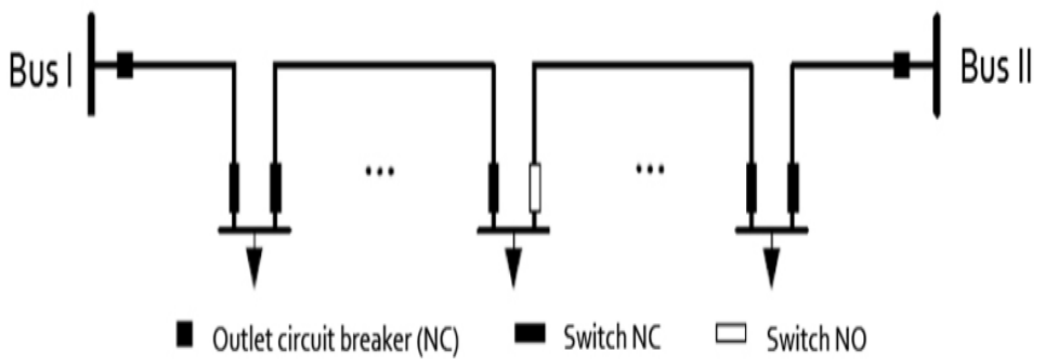

Fig. 2- 11 A 10KV double loop network

The typical grid structures of overhead lines include overhead multi-segmentation and single connection, and overhead multi-segmentation and moderate connection, as respectively shown in figures below:

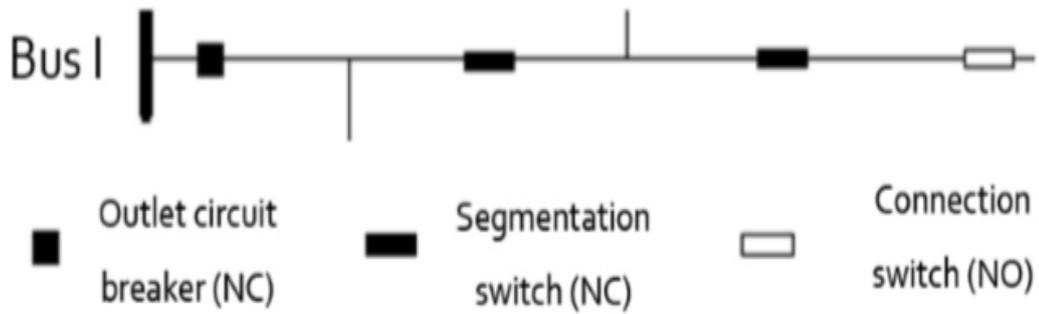

Fig. 2- 12 A 10KV multi segment and single connection shape

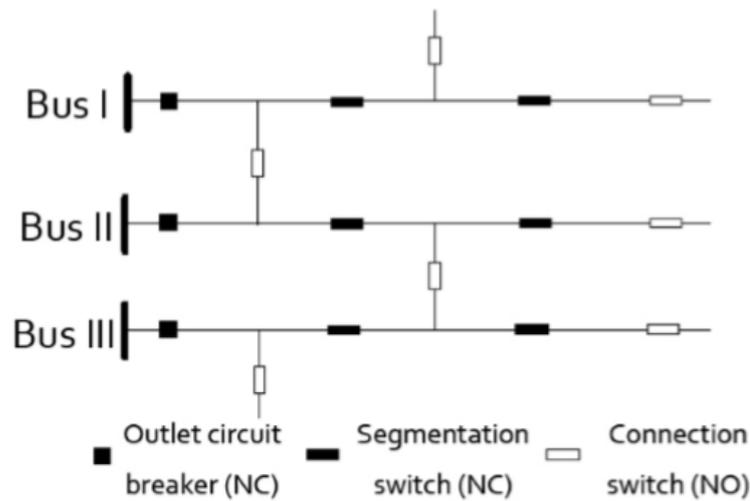

Fig. 2- 13 A schematic 10 KV diagram of moderate connection and multi segment structure

In China, investments in distribution networks have been continuously increased. The substation capacity and line length of the distribution network have been doubled. The grid structure tends to be rationalized and the power supply capability has been greatly strengthened, which have significant effects on the rapid growth of the urban and rural economic and social development. The intelligent micro grid construction in China is promoted steadily in order to adapt to rapid development of the distributed generation. It is mainly based on the local intelligent energy comprehensive utilization system built by local distribution grids with mutual compensation among various energy resources including wind, photovoltaic and natural gas, and power-grid-load coordinated interaction [71]. With high penetration level of renewable energy, it can almost achieve the balance between local energy production and consumption through storage devices and optimal energy allocation, and can realize flexible interaction with, or relatively independent operation of, public grids as needed. However, the intelligent microgrid in China is still at the initial stage [72]. Some pilot projects for demonstration of microgrid projects can be roughly classified in three categories: microgrid in remote regions, island microgrid and urban microgrid.

### 2.2.3.1 Pilot Projects for Demonstration of Microgrid Projects in China Promoting Distribution Network

Distribution network management and their proper allocation will stay major task for the power distribution authorities in China. The proper distribution network management can have fruitful outcomes for the locational pricing. The microgrid and smart grid concept cannot be developed without the distribution network. To create a decentralize market structure for china; development of microgrid is very important [73]. To promote the microgrid projects, Chinese power development industry is hugely focused on building and promoting the pilot projects and their integration with Microgrids. These are categorizes into three parts.

### 2.2.3.1.1 Microgrid in Remote Region

In China, the population density is low and ecological environment is vulnerable in remote regions, so that expansion of the conventional grid is costly and fossil fuels based power generation is harmful to environment. However, renewable energy sources including wind and solar power in remote regions are abundant. Therefore, using local distributed renewable energy in independent microgrid is an appropriate solution for supplying power to the remote regions. In China, a batch of microgrid projects has been currently constructed in remote regions in Tibet, Qinghai, Xinjiang and Inner Mongolia, for the power supply in these regions [73-74]. Table 2-2 is a clear demonstration of remote regions microgrid projects.

### 2.2.3.1.2 Microgrid in Island Mode

In China, the diesel based power generation with limited time is applied in islands. There are still about one million households of coastal or island residents living with lack of electricity.

Considering the high cost and the difficulties in transporting diesel to islands, construction of island microgrid using distributed abundant renewable energy on islands is a preferred solution for the power supply [74]. In China, a batch of island microgrid demonstration projects have been constructed to carry out theoretical, technical and application researches in practice. Table 2-3 demonstrates the Island mode microgrid projects in China.

### 2.2.3.1.3 Microgrid in Urban Region

In addition to microgrid in remote regions and island microgrids, China has also constructed urban microgrid demonstration projects. The key objectives of the demonstrations are to integrate distributed renewable energy, provide high quality and various reliable power supply services, and achieve comprehensive utilization of cooling and heating power.

In addition, there are some microgrid demonstration projects with special functions, such as the Sea Water Desalination Microgrid Project in Dafeng, Jiangsu. China has provided a flexible and comprehensive utilization path for integrating more distributed renewable energy through continuously increased investments in grid intelligence and multi-type energy complementation like wind and solar power complementation, and hydro and solar power complementation. The intelligent construction also becomes an important goal of grid development at all levels in China [75]. Table 2-4 depicts the urban area microgrid projects in China (table is continued to the next page).

Table 2- 2 Demonstration Projects of Microgrid in Remote Area of China

| Name/Location | Composition of System | Major Characteristics and Function |
|---|---|---|
| Shiquan River Microgrid in Ngari Prefecture of Tibet | 10MW photovoltaic power station, 6.4MW hydro power station, 10MW diesel generation unit and energy storage system | Multi-energy complementation between photovoltaic power, hydro power and thermal power; high altitude and awful weather |
| Jijiao Village Microgrid in Shigatse Prefecture of Tibet | Total installed capacity of 1.4MW, consisting of hydro power, photovoltaic power generation, wind power, battery energy storage and emergency diesel generation | Wind and solar power complementation; high altitude and difficult natural conditions |
| Rting Rngul Bon Dgon Microgrid in Nagqu Prefecture of Tibet | 15kW wind power, 6kW photovoltaic power generation and energy storage system | Wind and Solar power complementation; the first village microgrid in Tibet |
| 10MW Hydro and Photovoltaic Power Complementation Microgrid in Batang Town, Yushu County, Yushu Prefecture, Qinghai | 2MW single axis tracking photovoltaic power generation, 12.8MW hydro power and 15.2MW energy storage system | MW-level hydro and photovoltaic power complementation, one of largest photovoltaic microgrid power stations in China |
| Large Scale Photovoltaic Energy Storage Microgrid in Zadoi County, Yushu Prefecture, Qinghai | 3MW photovoltaic power generation and 3MW/12MWh two-way energy storage system | Parallel connection of multiple energy storage converters, and photovoltaic power and storage complementation and coordination control |

| Intelligent Photovoltaic Power and Storage Street Lamp Microgrid in Menyuan County, Haibei Prefecture, Qinghai | Centralized photovoltaic power generation and lithium battery energy storage | First similar system in plateau farming regions, changing the current situation where the lifetime of outdoor lead acid batteries is two years |
|---|---|---|
| New Energy Microgrid Demonstration Area in New Town of Turpan, Xinjiang | 13.4MW photovoltaic capacity (including photovoltaic and solar power) and energy storage system | Solar energy utilization and building integration project with the largest scale and most comprehensive technology application currently in China |
| Microgrid of Taiping Forest Farm in Eerguna, Inner Mongolia | 200kW photovoltaic power generation, 20kW wind power, 80kW diesel generation and 100kWh lead acid battery | Renewable energy power supply solution in forest farm in remote region |
| Old Barag Banner Microgrid in Hulunbuir, Inner Mongolia | 100kW photovoltaic power generation, 75kW wind power and 25kW×2h energy storage | Newly built migrant village and grid-connected microgrid |

Table 2- 3  Islanded Microgrid Projects in China (table continued to the next page)

| Name/Location | Composition of System | Major Characteristics and Function |
|---|---|---|
| Dong'ao Island MW-level Intelligent Microgrid in Zhuhai, Guangdong | 1MW photovoltaic power generation, 50kW wind power generation and 2MWh lead acid battery | Forming intelligent microgrid with diesel generator and transmission and distribution system, increasing the rate of renewable energy on the whole island to be over 70% |
| Dan'gan Island Microgrid in Zhuhai, Guangdong | 5kW photovoltaic power generation, 90kW wind power generation, 100kW diesel generation, 10kW wave power generation and 442kWh energy storage system | Having the first independent renewable energy power station in China; capable of utilizing wave energy; with capacity of seal water desalination at 60t/day |
| East Fushan Island Microgrid in Zhejiang | 100kW photovoltaic power generation, 210kW wind power generation, 200kW diesel generation and 1MWh lead acid battery energy storage system | An island with residents at the most east end of China; with capacity of seal water desalination at 50t/day |
| Nanji Island Microgrid in Zhejiang | 545kW photovoltaic power generation, 1MW wind power generation, 1MW diesel generation, 30kW ocean energy generation and 1MWh lead acid | Capable of utilizing ocean energy; introducing electric vehicle charging and changing station, intelligent electric energy meter, user interaction |

| | battery energy storage system | and other related advanced technologies |
|---|---|---|
| Luxi Island Microgrid in Zhejiang | 300kW photovoltaic power generation, 1.56MW wind power generation, 1.2MW diesel generation, 4MWh lead acid battery energy storage system and 500kW×15s super-capacitor energy storage | With function of flexible switching between microgrid connection and off-grid mode |
| Yongxing Island Microgrid in Sansha, Hainan | 500kW photovoltaic power generation and 1MWh lithium iron phosphate battery | The southernmost microgrid in China |

Table 2- 4  Microgrid Projects Demonstration in Urban Region of China (table continued to next page)

| Name/Location | Composition of System | Major Characteristics and Function |
|---|---|---|
| Integrated Microgrid at #2 Energy Station in Sino-Singapore Tianjin Eco-City | 400kW photovoltaic power generation, 1,489kW gas generation, 300kWh energy storage system, 2,340kW ground source heat pump unit and 1,636kW electric refrigeration unit | Flexible operation modes; coordinating comprehensive utilization of power, cooling and heating |
| Microgrid at Public House Exhibition Center in SinoSingapore Tianjin Eco-City | 300kW photovoltaic power generation, 648kWh lithium ion battery energy storage system and 2×50kW×60s super-capacitor energy storage system | "Zero energy consumption" building, overall balance in generating capacity and electricity consumption in the whole year |
| Nanjing Power Supply Company Microgrid in Jiangsu | 50kW photovoltaic power generation, 15kW wind power generation and 50kW lead acid battery energy storage system | Energy storage system is capable of smoothening fluctuation of wind and solar power output; enabling seamless switching of grid-connection/off-grid modes |
| Narada Power Source Co., Ltd. Microgrid in Zhejiang | 55kW photovoltaic power generation, 1.92MWh lead acid battery/lithium battery energy storage system and 100kW×60s super-capacitor energy storage | Battery energy storage is mainly used for "peak load shifting"; applying container type, with function modularization, realizing plug and play |

| Ecological Village Microgrid in Chengde, Hebei and Microgrid in Foshan, Guangdong | 50kW photovoltaic power generation, 60kW wind power generation, 128kWh lithium battery energy storage system and three 300kW gas turbines | Providing power supply guarantee for peasant households in the region, realizing double-power supply, and improving utilization voltage quality and combined cooling, heating and power technology |
|---|---|---|
| Intelligent Microgrid in Yanqing District, Beijing | 1.8MW photovoltaic power generation, 60kW wind power generation and 3.7MWh energy storage system | Combining distribution network structure design in China, with multistage microgrid framework and level-tolevel management, smoothly realizing grid-connection/off-grid switching |
| Photovoltaic power and Heat Storage Integration Microgrid of State Grid Hebei Electric Power Research Institute | 190kW PV power generation, 250kWh lithium iron phosphate battery energy storage system, 100kWh super-capacitor energy storage, EV charging pile and ground source heat pump | Connected with ground source heat pump to solve the problem of start-up impact; AC-DC hybrid microgrid |

Guangdong, Beijing, Fujian, Guizhou and other places in China; the key technologies of active distribution network and the optimization technology to promote renewable energy consumption have been developed rapidly; the planning and design method, operation management mode, source network load coordination and control technology, and other multidimensional innovation of active distribution network have greatly promoted the flexible local consumption of renewable energy and made the active distribution network to provide users with reliable, high-quality, efficient and diversified power services in a better manner. In addition, the rapid development of the internet of energy, the future power distribution system shall be oriented to the evolution of the internet of energy; the distribution network needs to promote the coordinated operation of "source-grid-load storage", integrate the coordinated multi-energy complementation of electricity, heat, gas, transportation, etc., promote the combination of distribution network and big data information system, and advance energy production and consumption "Internet +" business mode. However, the combination of distribution network and internet of energy is still in the stage of theoretical research, demonstration and trial, which needs further industrialized development [76].

## 2.2.5 Current Methods of Electrical Distribution Network Pricing in China

One of the core contents of the price formation mechanism is to determine whether prices are formed by market competition or government pricing. Because the scale economy of the power grid determines its natural monopoly characteristics, the price of transmission and distribution should be regulated by the government. Its price control should generally follow the following principles:

(1) Promote social distribution efficiency. If there is no external restraint mechanism for distribution monopolies, they will become market price makers instead of receivers. They may transform a portion of consumer surplus into producer surplus by setting monopoly prices, thereby distorting distribution efficiency. This requires the government to implement price controls to promote social distribution efficiency.

(2) Promote the improvement of efficiency and productivity. Through price regulation, enterprises are encouraged to optimize production and operation management, make full use of economies of scale, and constantly carry out technological and management innovations in an effort to achieve maximum production efficiency.

(3) Maintain the development potential of the enterprise. The power transmission and distribution industry has the characteristics of large investment and long investment recovery period, and it needs to advance the economic development speed. To this end, power transmission and distribution enterprises need to continuously make large-scale investments to ensure stable power transmission and supply, which requires the establishment of regulated prices At the same time, considering the ability of enterprises to self-accumulate and make large-scale investment continuously.

The following will make a detailed study on the price control methods and models of typical distribution networks to provide reference for China's incremental distribution price control.

### 2.2.5.1 Bidding Pricing Method

The bidding pricing law emphasizes that through the form of bidding, multiple companies can compete for franchise rights in an industry or business field. Under certain quality requirements, the companies that provide the lowest price can obtain the franchise rights. After the public bidding, the full market competition of the bidding entities enables the franchise to be given to the most efficient investment entities. At the same time, bidding entities should also make commitments such as investment scale, power distribution capacity, power supply reliability, service quality, and line loss rate. Relevant government authorities supervise and evaluate the power supply service standards agreed in the contract. If the agreed standards are not met, the price will be reduced accordingly.

Incremental power distribution business belongs to both public utilities and infrastructure, as well as the government's use of competitive methods (marketing mechanisms such as tendering) to authorize relevant organizations, sign agreements, clarify rights and responsibilities, and provide public products and services in a certain period of time. Therefore, most domestic incremental distribution networks currently adopt franchise models. In fact, the bidding pricing method for incremental distribution networks should be strictly called franchise bidding. In France, distribution companies only have the right to operate the distribution network and do not have distribution assets. The "concession contract" signed by the distribution company stipulates power supply reliability standards and punishment mechanisms, restrictions on various operating indicators, and environmental protection. Terms, value-added service categories, and other details, and the "Concession Contract" is not indefinite, and its operating rights are only 30 years. Therefore, it has improved the enthusiasm of power distribution companies to reduce costs, prices, and improve power quality.

## 2.2.5.2 Allowed or Permitted Income Method

The permitted income method is a traditional method of supervising the private monopoly industry. Its basic idea is to determine its permitted income according to the cost of the regulated enterprise and the prescribed rate of return. The purpose is to protect the income demand of the enterprise and prevent distribution companies from using monopoly status to set high prices. The annual permitted income of a power distribution enterprise includes the permitted cost, permitted income, and in-price tax. The formula is as follows:

$$F = C + Y + J \qquad （2\text{-}1）$$

$$Y = r \cdot B \qquad （2\text{-}2）$$

$$J = (C + Y) \cdot t \qquad （2\text{-}3）$$

Where

$F$ represents Annual permitted income for distribution companies

$C$ represents annual permitted costs, including annual operating expenses, annual depreciation charges, etc.

$Y$ Representing permitted income;

$J$ Representing the tax within the price;

$r$ For the return on investment;

$B$ An effective asset that can be accounted for

$t$ represents the VAT rate

Allowing income price supervision methods can enable enterprises to obtain reasonable returns, which is conducive to attracting investment. However, the incentive effect of distribution enterprises to improve service quality is not enough.

### 2.2.5.3 Ceiling Pricing Method

The ceiling price method means that the regulatory agency neither directly sets the price nor limits the profits, but only stipulates the maximum price or income. Since the mid-1980s, it has been adopted by Britain, Australia and other countries in infrastructure and utilities, and has evolved into a variety of models, such as the income ceiling method, which is now collectively known as the "RPI-X" model. Within the maximum level stipulated, the enterprise can obtain more benefits by reducing costs as much as possible. "RPI-X" model can be expressed as shown below:

$$P_t = P_{t-1}\left(1 + RPI - X\right) \pm Z \qquad (2\text{-}4)$$

Where

$P_t$ The distribution supervision price of the distribution company for the first year;

$P_{t-1}$ The electricity distribution supervision price for the first year of the distribution company;

$RPI$ For the retail price index, the inflation rate;

$X$ For the efficiency factor of the enterprise, it indicates that the requirements of the regulatory agency for improving the efficiency of the enterprise are the minimum level of production efficiency that the enterprise must meet;

$Z$ The adjustments are affected by external factors

### 2.2.5.4 Scale Competition Method

The basic principle of the scale competition law is based on the comparison of relative efficiency, which promotes competition between originally independent monopoly enterprises. Under the rule competition mode, the price of a power distribution company is determined by the industry's leading benchmark industry or the industry average, and has nothing to do with the company's own price. Therefore, in order to obtain more revenue, the power distribution company needs to improve efficiency and service quality strives to reduce costs. The regulatory model based on the scale competition method is as follows:

$$P_x = k_x C_x + (1 - k_x) C_m \tag{2-5}$$

$$P_x = k_x C_x + (1 - k_x) \sum_{y=1}^{n} (f_y C_y) \tag{2-6}$$

Where

$P_x$ The price of the distribution company under the scale competition mode

$k_x$ Representing the proportion of distribution equipment's own costs in the formation of distribution prices

$C_x$ For the cost of the distribution company itself;

$C_m$ Indicates the cost of the benchmark enterprise;

$f_y$ the proportion of the cost of the same type of enterprise in the formation of distribution prices; the cost of the same type of enterprise;

$n$ represents the number of companies of the same type.

Finally, according to the different characteristics of the distribution network, different pricing strategies are formulated. In an incremental distribution network with a large area and a relatively concentrated load, the problem of guiding the user to select a site should be considered, and it should be combined on the basis of the distribution strengthening model. The long run incremental cost (LRIC) model, under the condition that the permitted income is kept constant, gives the incremental power distribution enterprise a certain degree of independent pricing power. It can divide the price zone according to the long-term incremental cost of each node measured in the distribution network to guide users at low incremental cost; the location of the corresponding price zone is to guide users in high incremental cost price zones with high line utilization. For incremental distribution networks with smaller areas and relatively distributed loads, the location information is not required to be reflected. For relatively mature and stable loads, the prefecture-level stock distribution network with a complicated grid structure is inconvenient to calculate the network flow through the LRIC model, and only the distribution strengthening model can be used for pricing [77].

# Chapter 3 Research Methodology and System Modeling

This research methodology section has further divided into two sections. The first section of this study presented the power flow for LRIC and analyzes Network security for long run incremental cost (LRIC) and their impacts on the nodal injection in the distribution network on the basis of that power flow.

The power flow has been divided into three approaches. The base case flow approach and contingency approach broadly discusses and calculated the LRIC network security and its impact on pricing before and after nodal injection. For this purpose, Security factor has been calculated by formulation and compared on nodal injection based.

The second section presented the deep reinforcement learning algorithm (DRL) for calculation of distribution network pricing on the basis of security factor under the principle of LRIC. The DRL method optimizes the reactive power in a network flow to calculate and minimize the distribution network pricing.

## 3.1 Analysis of Power Flow Process Resulting LRIC Pricing Methodology

The power flow methodology determines the network utilization under N-1 network contingency state and normal arrangement. The injection of nodal generation or load causes great effect in utilization of the network at each single node.

The power flow analyze this change and divided into further three main parts i.e. Base case flow analysis, Contingency analysis, and incremental flow analysis by using iterative approach. This research is mainly focused on network security and network security factor calculation deriving distribution network pricing.

The network flow process is shown in the figure 3-1 (flowchart). The network security determination will helps to calculate the network security factor as a result efficient LRIC pricing can be found after getting proper output flows from the network.

## 3.2 Network Security and Charges Calculation Considering Network Security Factor

Network security is a term that is used to describe all power system aspects but it also play major role in the distribution network. Due to the continuous increase in the demand of electricity, the power networks in UK and all over the world are increasing the generation to meet the consumers demand. The incremental generation means to inject more and more generation points to the distribution networks to balance to create balance in the power system [78].

In doing so, the network security arises as a major challenging threat for the distribution companies and operators that affect the long run incremental and marginal cost. The network security creates N-1 contingency analysis in the distribution network system which greatly cause imbalance in the network distribution pricing [79].

Long run incremental cost pricing is the most advance pricing technique that is facing the network security challenge. This study analyzes, identifies this challenge first and formulated the network security factor. To improve and calculate the network pricing, network security needs to withstand all the credible network contingencies.

This study greatly contributes and proposes an improved approach for LRIC methodology prices on the basis of network security applying an IEEE-14 bus system in the network. The structure of IEEE-14 bus system can be seen in Appendix A (figure A-1).

To maintain the network security in an efficient manner, this study proposes a complete flowchart to overcome the network security challenge facing by LRIC methodology [80]. Therefore, this research produces a power flow analysis process that mainly consists of three aspects for network model under normal running condition.

1) Base case flow approach
2) Contingency analysis approach
3) Iterative approach (Incremental flow analysis)

These three approaches contribute towards the calculation of outputs from power flow analysis. Base case flow and contingency analysis contribute towards the calculation of security factor.

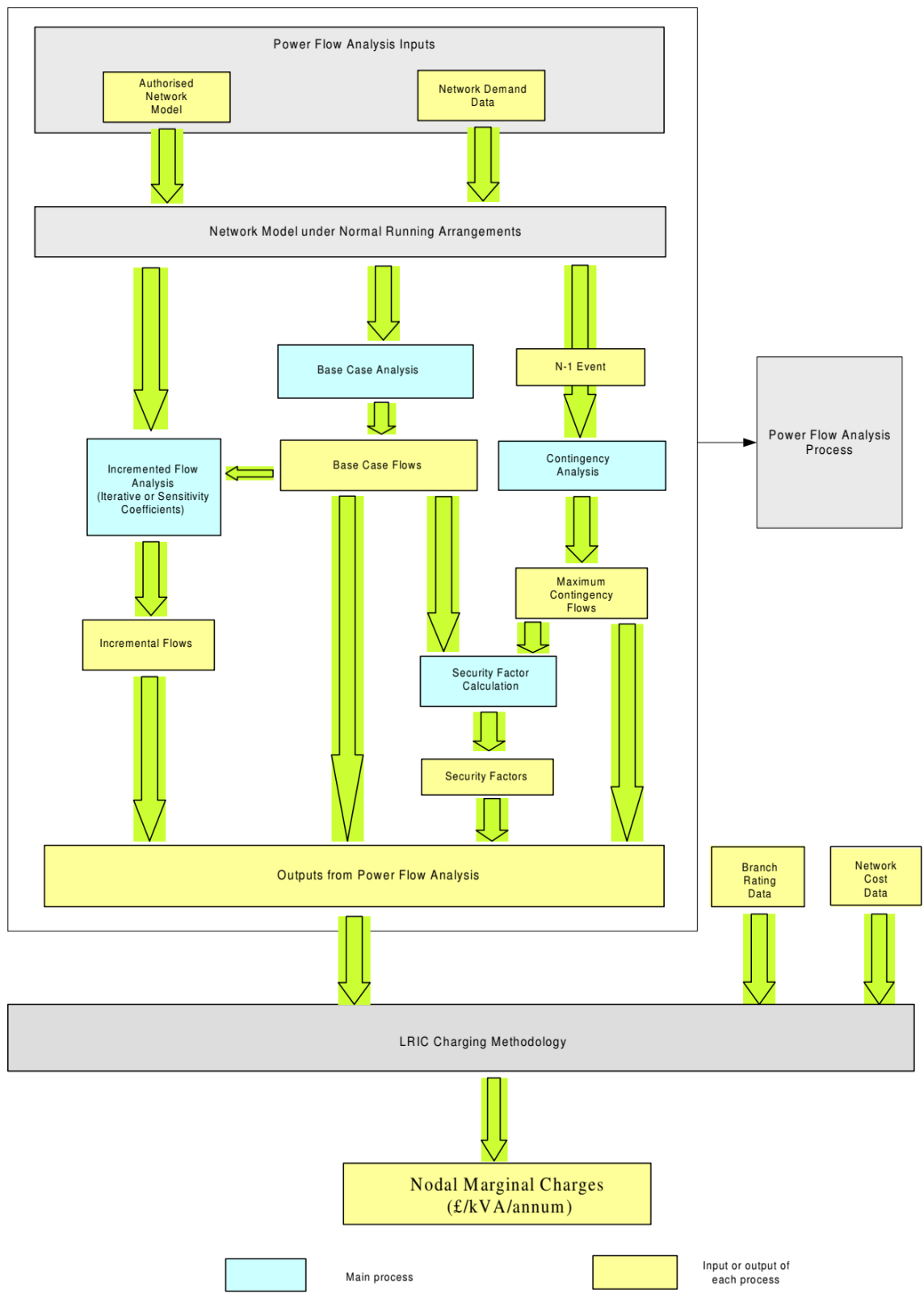

Fig. 3- 1 A flow chart of calculating LRIC Pricing under different analysis

## 3.3 Base Case Approach

The Base Case approach is required to determine the utilization of the network under Normal running condition. The key inputs to the Base Case method are the authorized Network Model and Network demand data. The output of the Base Case Analysis is the Base Case Flow in each Branch of the authorized network model.

## 3.4 Contingency Analysis Approach

This subject evaluates the variation in the utilization of network assets on behalf of Base Case approach. It is required to calculate the network utilization where the assets of network provide security of supply under N-1 contingencies. The authorized network, N-1 contingencies and demand data of a network are the key inputs for this approach. Each N-1 event will be allotted for every N-1 contingencies to meet the network security and supply. Finally, the output of this approach is the maximum contingency flow of the authorize network [81].

## 3.5 Calculation of Security Factor

From the flow chart in figure 3-1, security factor can be determined by the power flow analysis process. Security factors represent the change in utilization of a Branch between normal running arrangements and worst case N-1 Contingency conditions. The inputs playing role in security factor calculation are the base case flow approach and contingency analysis approach [82]. In general, it is formulated as:

$$Security\ Factor_B = \frac{Maximum\ Contingency\ Flow_B}{Base\ Case\ Flow_B})\qquad(3\text{-}1)$$

Here contingency flow is the maximum contingency flow at branch B. To calculate the security factor for N-1 contingent situation on a specific branch or line during the nodal injections into the network, it is represented in the following flow charts figure (3-2).

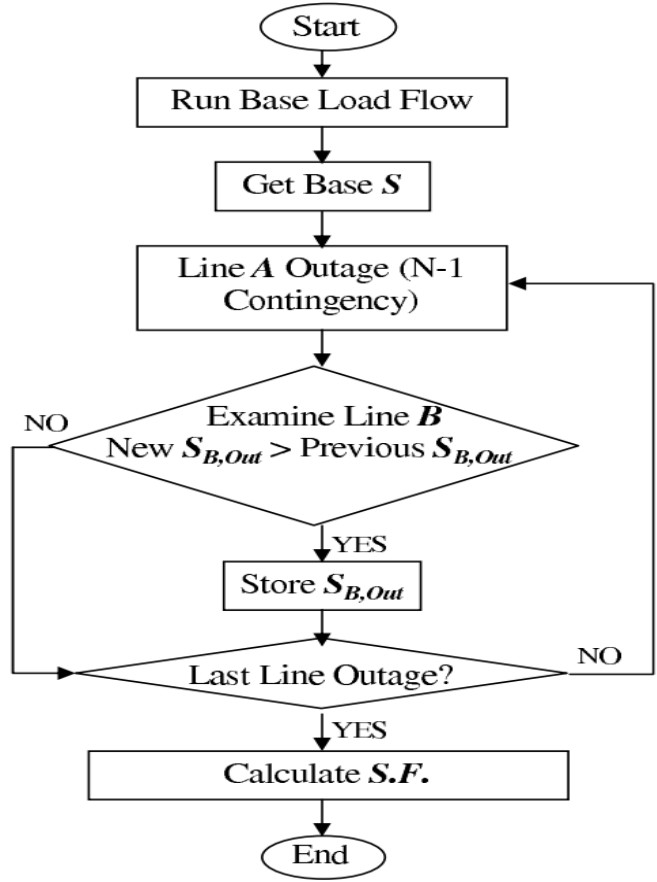

Fig. 3- 2 A stepwise flow chart for calculating Security factor in a network

So the security factor as Branch B, Base S component can be defined as:

$$S.F\left(S_i = \frac{P_{i,N-1}}{P_{i,original}}\right) \tag{3-2}$$

Where $P_{i,N-1}$ is the maximum contingency power flow as defined in (3-1) for Branch B now nominated as $B_i$ or $S_i$ or branch $i$ in N-1 contingency condition and $P_{i,original}$ is the power flow under normal condition.

Security factor provides enough spare capacity to withstand the network contingency in critical situation. The spare capacity is merged with maximum allowed loading level (MALL) which is defined as the spare capacity "$C$" defined by the networks security factor"$S.F$".

$$MALL = \frac{C}{S.F} \tag{3-3}$$

It is proved that for the lower value of maximum allowed loading level (MALL), the spare capacity for a network will be much smaller [83]. So this will help to increase the reinforcement asset cost. Figure A-2 in Appendix section denotes the maximum allowed loading level assuming 14 branches of IEEE-14 bus system to calculate the security factor and compare its difference in the network without and with security factor. The figure A-2 in Appendix explains the network with S.F lags the network without S.F. The maximum allowed level gets higher at some buses due to high security factor. The table A-1 evaluate and compare three node with the highest rated capacity but different MALL under N-1 contingent condition (see table A-1, in Appendix).

Security factor is very beneficial to calculate the maximum power in the network component and its effect on the loading level such as maximum allowed loading level (MALL), the component "S" mentioned in the equation (3-2). Hence, the security factor didn't remain the same at different branches because of the nodal injection at different distribution network points [84]. Due to the different values of security factor in branches with and without injecting nodes, the security factor needs to calculate again for better results and differentiation for power distribution network companies. So the recalculated security factor can be determined as its maximum power flow under N-1 contingent situation which can refer to maximum contingency situation as follows;

$$S_i^* = \frac{P_{i\,max}^*}{P_{i\,original}^*} \tag{3-4}$$

The (*) shows the new values after recalculating the security factor $S_i$.

$P_{i\,max}^*$ denotes the maximum amount of power flow under network N-1 contingent condition.

$P_{i\,original}^*$ is the branch power flow $i$.

$$S_i = \frac{P_{i\,max}}{P_{i\,original}} \tag{3-5}$$

The two equations (3-4) and (3-5) are not the same. Equation (3-4) demonstrates the security factor for Branch $i$ without any nodal injection into the network under N-1 contingent and normal network condition. While equation (3-5) determine the security factor after the injection of new nodes into the network that shows maximum flow through the network in N-1 contingent state at branch $i$. Therefore it is obvious that this network nodal injection creates network reinforcement in the network of distribution network operators [85-86]. So it is necessary to calculate the network horizon reinforcement for both cases defined in the equation (3-4) and equation (3-5).

Network reinforcement before the nodal injection can be formulated as;

$$T_i = \frac{\log\left(\frac{C_i}{S_i}\right) - \log(P_i)}{\log(1+d_i)} \tag{3-6}$$

Network reinforcement after the nodal injection can be formulated as;

$$T_i^* = \frac{\log\left(\frac{C_i}{S_i^*}\right) - \log(P_i^*)}{\log(1 + d_i)} \tag{3-7}$$

## 3.5.1 Case Study of IEEE 14- Bus System for Security Factor

The structure of IEEE-14 bus system is as given as;

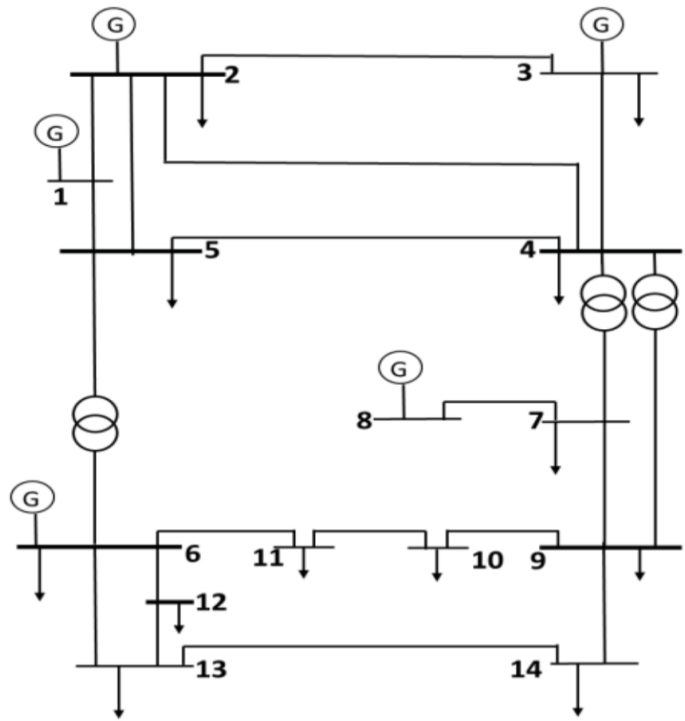

Fig. 3- 3 IEEE-14 Bus System

This study chooses IEEE-14 bus system for calculating the LRIC pricing network security. The IEEE-14 bus system consists of 14 buses and 20 nodes. The reinforcement horizon cost is assumed for this study on each branch is ¥800 million Yuan (RMB). I choose to take RMB currency here because this study is based on advice and suggestion for China's distribution network pricing in the long term from UK. In the circuit, the loading growth rate is 1% with the discount rate "d" is 0.03. A 10MW load is injected into the circuit at node number 14 of IEEE-14 bus system. Based on the assumption, this study inquires two conditions.

### 3.5.1.1 Security Factor Investigation on Each Branch

On the basis of mentioned assumed values in IEEE-14 bus system and equations derived for calculation of security factor, the security factor of each branch has calculated before and after the nodal injection. The detailed table of calculated security factor before and after the injecting node is shown in chapter 4 of this study (see table 4-1, chapter 4). The results are shown in figure 3-4. This figure depicts the differences in the security factor on different nodes before and after the injection at the nodes. It is noted that security factor is much higher at branch 6, 18 and 19 as compared to other branches in the network structure [87]. In case of LRIC condition, the prices without security factor can be lesser but it will not be cost reflective. The figure 3-4 shows the difference of security factor on each node and 20 branches in the IEEE-14 bus system.

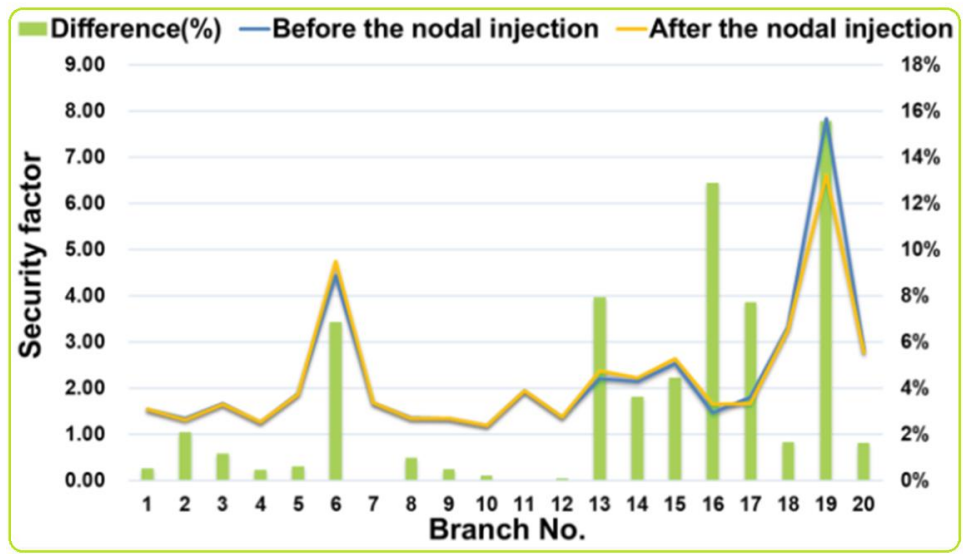

Fig. 3- 4 Comparison of Security Factor on different nodes

The higher security factor on some branches results in lower utilization capacity under normal conditions. It can be seen that at branch 2, the differences is a bit higher because of its non-functioning and as a result the adjacent node 3 is responsible for carrying that non-functioning load that results in increase in the security factor at node 6 [88]. In chapter 4 Table (4-2) explains the results on the branches (6, 18 and 19) and their higher security factor due to the maximum normal flow, contingency flow.

### 3.5.1.2 LRIC Comparison Before and After the Nodal Injection

The LRIC pricing is greatly affected by the network security and security factor. Indeed, it has been studied, analysed; and the purpose of this research is to resolve the network security issue. This concept put efforts to compare the LRIC pricing before finding network security factor and after finding network security factor. As it has been proved that without network security, a distribution network cannot withstand the N-1 contingencies. So after injecting a nodal generation into the network, a different new LRIC charging has been produced including network security withstanding network contingencies. The difference between the LRIC pricing with and without network security is that; the LRIC pricing with network security is no more prone to the network contingencies [89]. The previous LRIC pricing without network security (without nodal addition) and the new LRIC (with nodal addition) achieved by this study can be shown in the figure (3-5).

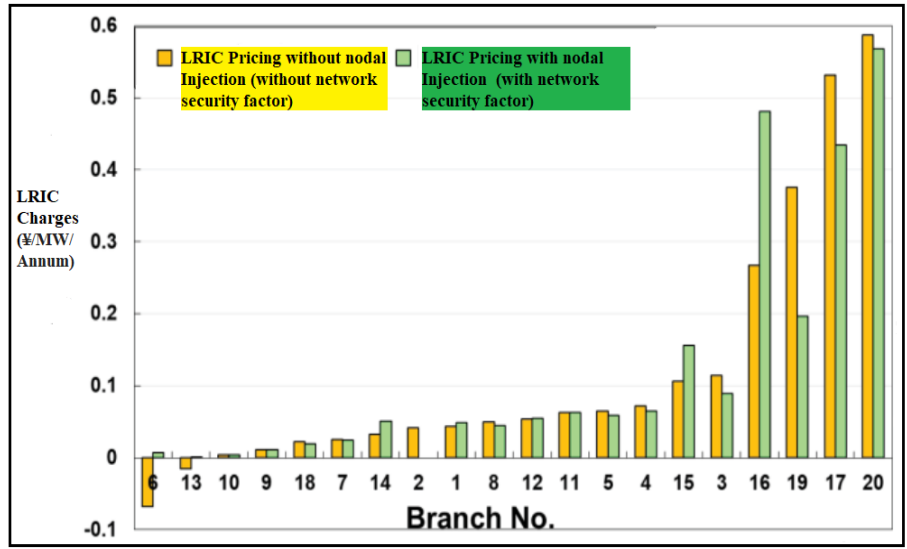

Fig. 3- 5 Comparison of LRIC Pricing with and without nodal injection

The new LRIC pricing is much more beneficial to get the fairer pricing in the distribution network to supply cheap electricity distribution for the consumers.

### 3.5.1.3 LRIC Charges at different Nodes

The principle of LRIC pricing method is to calculate the incremental nodal charges, which represent the brought forward reinforcement costs caused by the addition of an increment of demand or generation at each network node. The LRIC charges difference is a positive point for the distribution network companies because this difference reflects the difference in distribution network costs which is responsible for providing economic signals. These different charges are drawn in chapter 4 (figure 4-1). From the figure it can be seen that LRIC charges increase as it is calculated in ascending order from node 4 to node 14. But the charges at node 14 is the most higher and at node 5 is the lowest of them. This is because the node 14 is far away from the input generation point that requires more networks to transport power to it as a result this node is charged with higher network costs. Node 5 is closer to the generating point hence it has lower distribution network cost [90].

On the other hand, the additions of increment at each network node also have influence on results. Figure (4-2) in chapter 4 presents the results of LRIC pricing impacts by means of different nodal injection values. From the figure it is suggested that smaller node injection is much more effective and beneficial for LRIC charging methodology.

## 3.6 Iterative Approach

The iterative approach calculates the incremental flows in the network by applying the increment to the respective node following base case approach and contingency analysis approach to produce nodal marginal charges for LRIC pricing methodology. The iterative approach is based on reactive and active paper.

This thesis proposed reactive power calculation for distribution network to get an efficient LRIC pricing while keeping the network security. This study has deeply concerns about the lack of attention towards the role of reactive power and its effect on distribution network pricing in the long term. This study presented at least three convincing reason about the reactive power calculation and optimization for efficient network security so that the prices calculation and balancing of the network gets easier for the distribution network operators (DNOs) [91]. The three main talking points about reactive power for distribution network are as follows:

> ➢ The free flow of reactive power in the distribution network can results in increase in losses; due to which the network security can't withstand any more causing N-1 contingent situation.
> ➢ The reactive power has major influence on voltage flow and voltage deviation in the distribution resulting immediate increase or decrease voltage in the respective network system.
> ➢ Capacity utilization is one of the factor as reactive power utilizes both distribution and transmission network capacity.

To secure efficient network security and LRIC pricing, reactive power optimization and its reduction is necessary in the network to decrease the distribution network pricing for current and future consumers. This research make "Reactive Power optimization in the network" as the bold statement to solve the LRIC network security challenge and reduce the distribution network pricing by regulating and minimizing the reactive power values in the network [92-93].

In doing so, this study presented concept and methodology called Deep Reinforcement Learning (DRL) method to optimize the reactive power values in the network. To apply this method, an IEEE-14 bus system (see appendix A, figure A-1) is designed as its mathematical model and practically simulated this method in MATLAB programming software. The reference power of the system has kept at 100 MVA. The IEEE-14 bus node consists of five generators, four adjustable transformers and one reactive power compensation point. Deep reinforcement learning uses iteration phenomena for reactive power calculation by doing several times iteration following a power flow analysis for the defined values in the program and creates the result.

## 3.6.1 Significance of Deep Reinforcement Learning (DRL) and its Task for this Research

Deep reinforcement learning (DRL) is a new algorithm combining deep learning (DL) derived from neural network with reinforcement learning (RL) in machine learning [94-95].This method realizes learning from perception to decision making, which can extract and analyze environmental information to guide the physical model to make correct actions [96]. Through the input data, the deep neural network constructed by deep reinforcement learning (DRL) can process the data and output the action without manual intervention [97].

Deep reinforcement learning (DRL) has great advantages in dealing with complex and multifaceted problems because of its combination of perception and decision-making power. Most of the problems in the power system are dynamic, multi constrained and nonlinear [98]. The traditional calculation methods have different problems, such as poor convergence or low calculation accuracy. The method of deep reinforcement learning is the key to optimize the reactive power of the distribution power system has the characteristics of flexibility and good convergence, which can effectively control the voltage offset and regulate the distribution of reactive power [99-102].

This research aims to solve the problem of reactive power optimization of distribution network and their pricing by deep reinforcement learning. Firstly, the basic state data of distribution network is obtained; secondly, the neural network processing data is used to select the appropriate action output; then, the power flow calculation is used to verify whether the optimization is successful. The goal of the project is to use this model to calculate and maintain the network security while controlling the reactive power losses of the distribution network in a timely manner and maintain the stable operation of the distribution network on pricing perspectives [103-104]. The main tasks of the study are as follows:

1. Determine the mathematical model of reactive power optimization and select the appropriate deep reinforcement learning algorithm (DQN);

2. Design reactive power optimization process based on deep reinforcement learning along with IEEE 14-node model and power flow calculation tool in MATLAB.

3. Use MATLAB to establish deep reinforcement learning network and realize reactive power optimization.

4. Adjust parameters, optimize the model of reactive power optimization algorithm, train the model, and test the performance of the trained model.

### 3.6.2 Methodology for Calculating Reactive Power Based on DRL

The basic flow chart for calculating reactive power accordingly using Deep reinforcement learning (DRL) method is given below.

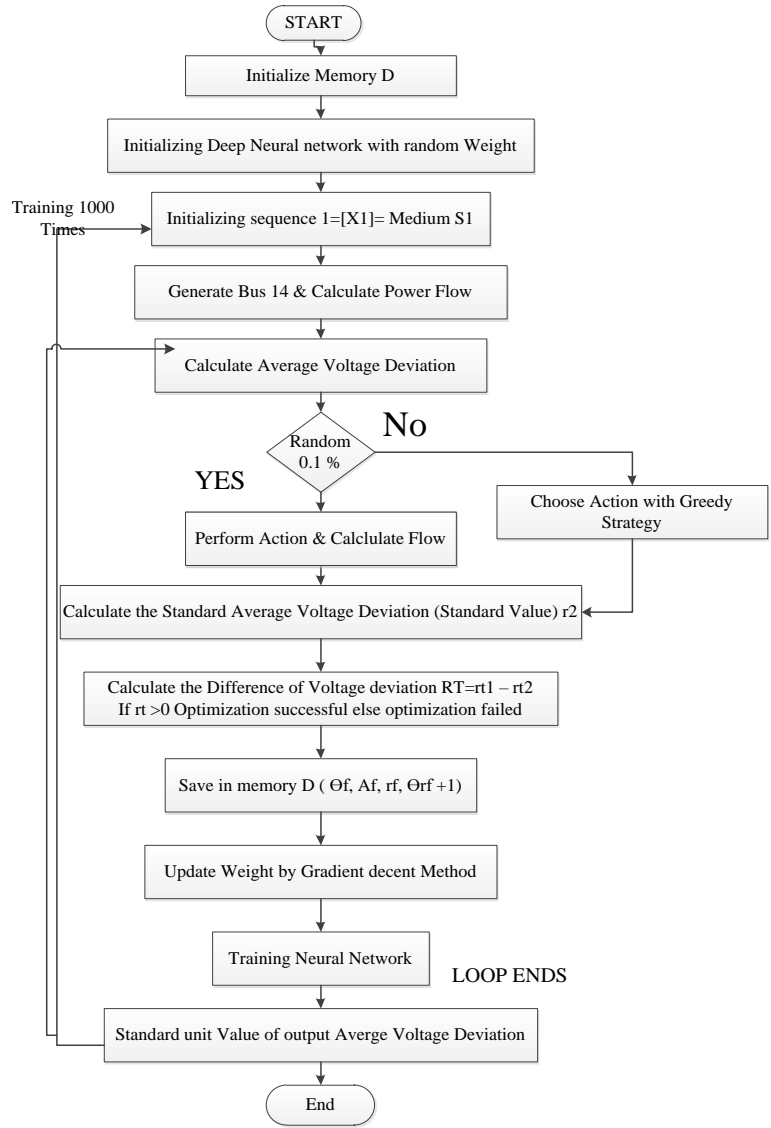

Fig. 3- 6 Flow chart of reactive power optimization based on DRL

After the deep neural network is established, the network is initialized and the power flow is calculated. Average voltage deviation is achieved, which is denoted by rt1. The power flow and voltage mean deviation rt2 are calculated again with the changed data after the action, if the voltage deviation after the action is smaller than before the action.

$$rt = rt1 - rt2 > 0 \qquad (3\text{-}8)$$

It shows that the action is effective and the optimization is successful. The bigger the RT is, the better the effect is and the better the optimization is. The random number Rand is used to represent the probability when selecting actions. When the Rand is less than 0.1, it is regarded as the occurrence of a small probability event. At this time, a random action will be selected from the action set. If the small probability event does not occur, the action with the largest Q value will be selected according to the greedy strategy through the calculation of neural network.

Regardless of the success or failure of optimization, data should be stored in the memory bank. The main function of the memory bank is experience playback, that is, uniform random sampling from the memory bank, whose capacity is set to 100. However, the training is usually thousands of times, so the pointer is added to the memory bank. Once the memory bank is full, the pointer will return to the first grid again, and the subsequent data will cover the stored data in order according to the position of the pointer.

16 sets of data were sampled uniformly and randomly from the memory bank, and the one with the largest TD target was used to perform the gradient descent algorithm, then the network weight was updated and the neural network was trained. To increase the speed of the run, set the weight update to update the weight every 20 steps during a study. The reason for this is that some optimized functions are very complicated, and sometimes the neuron output will appear some flat areas. The weight of the traditional neural network will change very little. The 20-step update of the weight can avoid the low efficiency of the algorithm and the long calculation time.

### 3.6.3 Training pseudo code for Distribution reinforcement learning

As discussed, Distribution reinforcement learning is a neural network training program based method, this study trained the neural network algorithm in MATLAB to get and optimize reactive power in many iteration to get the accurate real and reactive power results. The training code used in this research is presented in the following table 3-1 for distribution reinforcement learning algorithm. The whole process of optimizing reactive power is done through this MATLAB coding algorithm [105].

Table 3- 1 Neural Network Training pseudo code

**DQN Algorithm**

Initialize the memory base D, and the amount of data that can be accommodate is N

Initialization of Q-Value Functions with Random Weights loop $\theta$

Repeat

Initialize the first state of the event $s_1$ and get input through preprocessing loop

Repeat

Selecting a Random Action ε with Probability $a_t$

If no small probability event occurs, greedy strategy is used to select $a_t = \arg\max_a Q(\phi(s_t), a; \theta)$

Executive action a, Observation reward $r_t$ and image $x_{t+1}$

Set up $s_{t+1} = s_t, a_t, x_{t+1}$, Preprocessing $\phi_{t+1} = \phi(s_{t+1})$

In memory bank D , Save conversion in $(\phi_t, a_t, r_t, \phi_{t+1})$

In memory bank D, Conversion of sample data $(\phi_j, a_j, r_j, \phi_{j+1})$ Conduct uniform random sampling

Determine whether the event terminates, if so，TD Target is $r_j$ ；Otherwise, use TD Target network $\theta^-$ Computing TD targets $r + \gamma \max_{a'} Q(s', a'; \theta)$

Execute a gradient descent algorithm $\Delta\theta = \alpha[r + \gamma \max_{a'} Q(s', a'; \theta) - Q(s, a; \theta)]\nabla Q(s, a; \theta)$

Training Neural Network

End cycle

This table derives the training algorithm for optimizing reactive power in the distribution network and power grid by executing it in MATLAB simulation. The neural network training window is shown in the Appendix section, (Figure B-1).

### 3.6.4 Mathematical Modeling of Reactive Power Optimization

Mathematically, reactive power is calculated with different method. This mathematical modelling uses different algorithms and considers different constraints to formulate reactive power. Reactive power optimization of power system refers to that when the structure and load of distribution network system/power grid have been given, some control variables have been optimized to make one or more performance indexes of the system reach the optimal level under the premise of satisfying various constraints of the system [106-109].

#### 3.6.4.1 Classical Reactive Power Optimization model

The mathematical model of reactive power optimization is generally nonlinear, including objective function, equality constraint and inequality constraint.

$$\begin{cases} \min f(u,x) \\ s.t. g(u,x) = 0 \\ h(u,x) \le 0 \end{cases} \tag{3-9}$$

In the above formula, it is the control variable, the generator terminal voltage of non PQ node, the transformation ratio of adjustable transformer and the input number of reactive device. It is the state variable, including the node voltage of the power system and the reactive power of the line. $f(u,x)$ is the objective function of reactive power optimization, which is usually the network loss or voltage deviation value. $g(u,x)$ is a nodal power flow equation, which needs to satisfy certain constraints. $h(u,x)$ are inequality equations, including inequality equations for control variables and inequality equations for state variables.

#### 3.6.4.1.1 Objective Functions

#### 3.6.4.1.1.1 Aiming at Minimizing Network Loss

Taking the operation economy of power distribution network as the goal, the objective function is selected as the active power loss minimization

$$\min f_1 = \min P_{loss} = \min \sum_{k=1}^{n_l} G_{k(i,j)}[U_i^2 + U_j^2 - 2U_i U_j \cos(\theta_i - \theta_j)] \tag{3-10}$$

Among them, $n_l$ represents the total number of network branches; $G_{k(i,j)}$ represents the label of the first end of the grid is $i$; The conductance of the path labeled $j$ at the end; $U_i, U_j$ represents the voltage of nodes $i$ and $j$ respectively; $\theta_i, \theta_j$ represents the phase angle of node $i, j$.

### 3.6.4.1.1.2 Aiming to minimize the capacity of reactive compensation equipment

Taking the minimum investment of distribution network and power grid equipment as the goal, the objective function is selected as the minimum capacity of reactive compensation equipment.

$$\min f_2 = \min f_{invest} = \sum_{i \in q} C_i \left| Q_{qi} \right| \tag{3-11}$$

Here, $i$ represents the node that can adjust the reactive compensation capacity; $Q_{qi}$ represents the reactive capacity actually compensated by the node; $C_i$ represents the investment required to compensate the unit capacity, When $C_i = 1$, the objective of optimization is to minimize the amount of reactive compensation.

### 3.6.4.1.1.3 Aiming to minimize the voltage deviation

Voltage is an important aspect of reactive power in a power grid and in a distribution/transmission system as discussed in (Chapter 3, 3.6). Taking the network security of distribution network as the goal, the objective function is selected as the minimum voltage deviation

$$\min f_3 = \min f_{vd} = \min \sum_{i \in N_{PQ}} \left| \frac{U_i^{spec} - U_i}{U_i^{max} - U_i^{min}} \right| \tag{3-12}$$

Here, $N_{PQ}$ represents **PQ** bus set, $U_i^{spec}$ represents the voltage reference value of the nod, $U_i$ represents the voltage of the node $i$, $U_i^{min}$、$U_i^{max}$ represents the maximum and minimum voltage of node $i$ respectively.

### 3.6.4.1.1.4 Aiming at the maximum margin of voltage safety and stability

The static stability margin of power grid can be measured by the minimum singular value of Jacobean matrix of convergence power flow. The larger the margin is, the greater chance of stability in the network is expected. Taking the stability of power network and the local grid operation as the goal, the objective function is chosen as the maximum voltage static stability margin.

$$\min f_4 = \min \delta_0 = \min(\frac{1}{\delta_{min}}) \tag{3-13}$$

### 3.6.4.1.1.5 Multi-objective weighted reactive power optimization model

After normalization processing of the objective function in the classical reactive power optimization model above, a multi-objective weighted reactive power optimization model can be established through the option value:

$$\min F = \min(\lambda_1 f_1^* + \lambda_2 f_2^* + \lambda_3 f_3^* + \lambda_4 f_4^*) = \min(\lambda_1 p_{loss}^* + \lambda_2 p_{invest}^* + \lambda_3 v_{ad}^* + \lambda_4 \delta_0^*)$$
(3-14)

The normalization process is following as;

$$\begin{cases} p_{loss}^* = (p_{loss} - p_{loss.\min})(p_{loss.\max} - p_{loss.\min})^{-1} \\ p_{invest}^* = (p_{invest} - p_{invest.\min})(p_{invest.\max} - p_{invest.\min})^{-1} \\ v_{ad}^* = (v_{ad} - v_{ad.\min})(v_{ad.\max} - v_{ad.\min})^{-1} \\ \delta_0^* = (\delta_0 - \delta_{0.\min})(\delta_{0.\max} - \delta_{0.\min})^{-1} \end{cases}$$
(3-15)

### 3.6.4.1.2 Variable constraint equation

Inequality constraint of control variables are given as;

$$\begin{cases} U_{G,i}^{\min} \le U_{G,i} \le U_{G,i}^{\max}, i = 1, 2, \ldots, N_G \\ Q_{C,i}^{\min} \le Q_{C,i} \le Q_{C,i}^{\max}, i = 1, 2, \ldots, N_C \\ t_{T,i}^{\min} \le t_{T,i} \le t_{T,i}^{\max}, i = 1, 2, \ldots, N_T \end{cases}$$
(3-16)

In the above formula,

$N_G$ represents the number of generators, $N_C$ represents the number of shunt capacitors, and $N_T$ represents the number of transformers. $U_{G,i}$ represents the voltage of generator node $i$, $t_{T,i}$ represents the transformer ratio of node $i$, and $Q_{C,i}$ represents the reactive compensation capacity of node $i$. $U_{G,i}^{\min}$、$U_{G,i}^{\max}$、$Q_{C,i}^{\min}$、$Q_{C,i}^{\max}$、$t_{T,i}^{\min}$、$t_{T,i}^{\max}$ points represent the upper and lower limits of corresponding variables.

The inequality constraints of state variables are given as;

$$\begin{cases} U_i^{\min} \le U_i \le U_i^{\max}, i = 1, 2, \ldots, N \\ \theta_{ij}^{\min} \le \theta_{ij} \le \theta_{ij}^{\max}, j = 1, 2, \ldots, N \end{cases}$$
(3-17)

In this equation,

$\theta_{ij}$ is the voltage phase angle difference between node $i$ and $j$, $U_i$ is the voltage size of node $i$. $\theta_{ij}^{\min}$、 $\theta_{ij}^{\max}$, $U_i^{\min}$、$U_i^{\max}$ are the upper and lower limit values of corresponding variables respectively, and $N$ is the number of system nodes.

### 3.6.4.1.3 Power constraint equation

$$\begin{cases} P_i - U_i \sum_{j=1}^{N} U_j [G_{ij} \cos \theta_{ij} + B_{ij} \sin \theta_{ij}] = 0, i \in N_{PV}, N_{PQ} \\ Q_i - U_i \sum_{j=1}^{N} U_j [G_{ij} \sin \theta_{ij} - B_{ij} \cos \theta_{ij}] = 0, i \in N_{PQ} \end{cases} \quad (3-18)$$

In the above equation, $P_i$、 $Q_i$ are the injection amount of active and reactive power of node $i$ respectively, $G_{ij}$、 $B_{ij}$ are the conductance and susceptance between nodes $i$ and $j$ respectively.

### 3.6.4.2 Reactive power optimization in power market environment

In the electricity market environment, in order to ensure the economic benefits of power plants and distribution network, the reasonable pricing of reactive power is also very important. The objective function of the reactive power optimization model of the power market is to minimize the cost of active power loss and reactive power expenditure. The mathematical calculation for this statement can be shown in the equation here;

$$\min f = \lambda_{loss} p_{loss} + \sum_{k=1}^{Gs} f_{Qk}(Q_{Gk}) \quad (3-19)$$

Here, $\lambda_{loss}$ stands for the marginal price of active power; $p_{loss}$ stands for the marginal loss of active power; $f_{Qk}(Q_{Gk})$ stands for the cost of reactive power expenditure of the first non-grid company, which can be used to charge the reactive power of the generator in the form of segmented quotation. The constraints of power market are the same as those of classical reactive power optimization model.

### 3.6.4.2 Reactive power optimization including wind power system considering case function of China

When the distributed generation is connected to the grid, the voltage fluctuation often occurs. With more and more wind power connected to the grid and distribution network, the voltage fluctuation is more and more serious. This experience is both shared by UK and China to have the same challenge.

The increase of wind speed will lead to the bus voltage collapse. The change of system operation mode will also lead to the bus voltage fluctuation, and the grid voltage will collapse with a little carelessness. The key points of wind power reactive power optimization are: active power, reactive power, relationship between wind speed and load fluctuation, switching control and compensation capacity of reactive power compensation equipment. Due to the variable wind speed, a comprehensive index of reactive power optimization based on the probability of scenario occurrence is proposed, which is composed of active loss and static voltage stability margin.

$$\lambda = \omega_1 \lambda_1 + \omega_2 \frac{1}{\lambda_2} \tag{3-20}$$

$$\lambda_1 = \sum_{k=1}^{n} p_k p_{loss}^k, \lambda_2 = \sum_{k=1}^{n} p_k \delta_k \tag{3-21}$$

Here,

$\omega_1, \omega_2$ represent the weight of the two indicators; $\lambda_1, \lambda_2$ are the expected values of active loss and static voltage stability margin; $p_{loss}^k, \delta_k$ are the active loss and static voltage stability margin of the $k$ scenario respectively.

Next, taking transformer ratio and input number of reactive power compensation device as control variables, and taking the objective of minimizing the comprehensive index of reactive power optimization, the reactive power optimization model is established.

$$\begin{cases} \min \lambda \\ s.t. h(u, x, p_w^k) = 0, k = 1, 2, \ldots, n \\ g(u, x, p_w^k) \leq 0, k = 1, 2, \ldots, n \end{cases} \tag{3-22}$$

$u, x$ are control variables and state variables respectively, $h$ is equality constraint, i.e. power flow equation; $g$ is inequality constraint, $p_w^k$ is control variable constraint and the active power output of the wind turbine in the $k$ scenario.

# Chapter 4 Results Analysis and Discussions

This section of the thesis presented the results practically derived from the whole research and analyzes each and every result in detail. This research discusses security factor calculation results achieved and compare it with the nodal injection before and after the network. The effect of LRIC distribution pricing for different node has been compared achieving highest network security and lower network cost. The effect of voltage values and its deviation with respect to reactive power in the network is presented.

In last, the reactive power optimization is performed for several repetitive periods through different reactive power optimization technique proposed by this study to reduce the reactive power losses and its network security so as to efficiently decrease the distribution network pricing. The method of distribution reinforcement learning (DRL) algorithm is being used for optimization of reactive power in the distribution network with the help of IEEE14 bus node system in MATLAB simulation.

## 4.1 Results and Analysis of Security factor calculation applying IEEE-14 bus

The results of the table (4-1) and (4-2) are carried out from the section (3.5.1.1) that compares the two security factor values for 14 buses and 20 branches of IEEE-14 bus system. Security factor are compared on the basis of nodal injection generation at the branch 14 of the IEEE-14 bus system.

The results "from at each node" "to nodes" are generated for security factor calculated with and without the node injection. The results for 14 buses and 20 branches are given as;

Table 4- 1 Security Factor Comparison with and without Nodal injection

| Branch Number | From Node | To Node | Security Factor without node injection | Security Factor with node injection |
|:---:|:---:|:---:|:---:|:---:|
| 1 | 1 | 2 | 1.5361 | 1.5432 |
| 2 | 2 | 3 | 1.3427 | 1.3151 |
| 3 | 2 | 4 | 1.6713 | 1.6522 |
| 4 | 1 | 5 | 1.2643 | 1.2582 |
| 5 | 2 | 5 | 1.8789 | 1.8688 |
| 6 | 3 | 4 | 4.4322 | 4.7345 |
| 7 | 4 | 5 | 1.6822 | 1.6814 |
| 8 | 5 | 6 | 1.3524 | 1.3394 |
| 9 | 4 | 7 | 1.3341 | 1.3405 |
| 1 | 7 | 8 | 1.1807 | 1.1831 |
| 1 | 4 | 9 | 1.9340 | 1.9341 |
| 1 | 7 | 9 | 1.3699 | 1.3708 |
| 1 | 9 | 1 | 2.1970 | 2.3718 |
| 1 | 6 | 1 | 2.1433 | 2.2205 |
| 1 | 6 | 1 | 2.5233 | 2.6354 |
| 1 | 6 | 1 | 1.4602 | 1.6484 |
| 1 | 9 | 1 | 1.7994 | 3.2656 |
| 1 | 1 | 1 | 3.3207 | 1.6607 |
| 1 | 1 | 1 | 7.8346 | 6.6156 |
| 2 | 1 | 1 | 2.8160 | 2.7699 |

The N-1 contingency is observed according to the values of security factor obtained on different branches. The high security factor highlights the lesser capacity exploitation value functioning under normal state. The results are further explained in table 4-2.

Table 4- 2 The branches with high security factor

| Branch number | Branch outage flow | Normal flow | Contingency flow | Security factor |
|---------------|--------------------|-------------|------------------|-----------------|
| 6 | 2 | 0.233 | 1.031 | 4.4322 |
| 18 | 7 | 0.039 | 0.129 | 3.3207 |
| 19 | 16 | 0.017 | 0.132 | 7.8346 |

These results are the succeeding and shortlisted results from the table (4-1) in which the security factor is differentiated accordingly node addition. The three results at the three respective branches (6, 18, and 19) are found comparatively higher due to the maximum flow, maximum contingency flow and normal flow at these branches. So it is actually a concern for the distribution network due to their higher security factor.

## 4.2 Different LRIC Charges at Different Nodes

Section (3.5.1.3) from Chapter 3 depicted this result where different charges have been calculated at different node for LRIC methodology. It is shown in the figure (4-1).

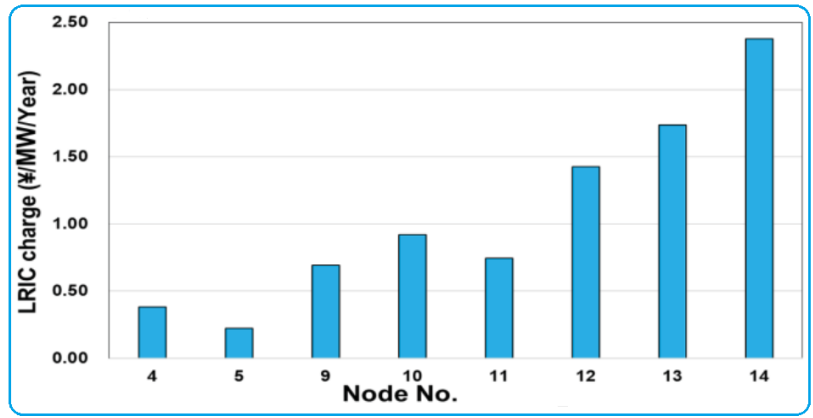

Fig. 4- 1 Different LRIC Charges for different nodes

The charges at node 14 is the most higher and at node 5 is the lowest of them. This is because the node 14 is far away from the input generation point that requires more networks to transport power to it as a result this node is charged with higher network costs. Node 5 is closer to the generating point hence it has lower distribution network cost. The more node closer to the input generation capacity, the lesser will be the network cost.

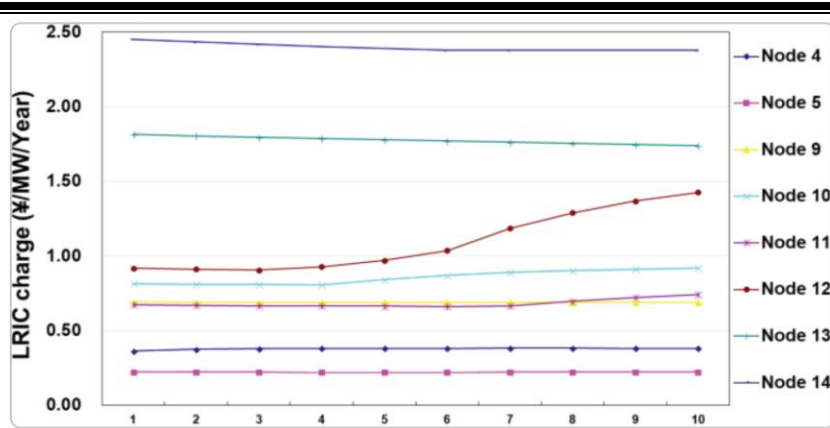

Fig. 4- 2  LRIC pricing with different node injection calculations

From these results as shown in the above figure, the nodal generation has influenced the output results. The node 14 stays the horizontal straight line at the top because of its high distribution cost followed by node 13. Node 12 is increased suddenly after an injection of further 12MW instead of 1MW. The more higher the injection values the highest peak will be found that indicates the increase in the distribution network pricing at the specific nodal point.

## 4.4 Reactive Power effect over voltage deviation in a distribution network

Section 3.6 of Chapter 3, already highlighted the importance of reactive power and its effect in the variation of voltage in distribution network considering the distribution network security in the longer term and its impact over the distribution network cost pricing. Voltage variation if not dealt properly; it can be found a major factor in the network contingency linking to network security dilemma.

This research methodology presents the novel approach of dealing with this issue by considering it a key major factor affecting the network contingency and network future reinforcement cost. So, this study calculated the voltage deviation with the help of MATLAB simulation.

The deep reinforcement learning (DRL) algorithm makes it possible to point out and sort out the voltage variation issue with the help of IEEE-14 bus system and uses several learning optimization to efficiently regulate and decrease the voltage deviation ratio in the network. This method not only makes it possible to regulate the voltage deviation in the distribution network; as well it provides a prompt solution of dealing voltage deviation for transmission network and power grid. The results are further supported by using Single learning and multiple learning techniques in MATLAB by using DRL.

### 4.4.1 Trend of Average Voltage in Single Learning

In deep reinforcement learning, the effect of single learning is very important. The following figure takes 35 optimizations in certain learning and uses the average voltage deviation as the vertical coordinate. It can be seen that the average voltage deviation after distribution network system/power grid optimization shows a downward trend with increase in optimization times.

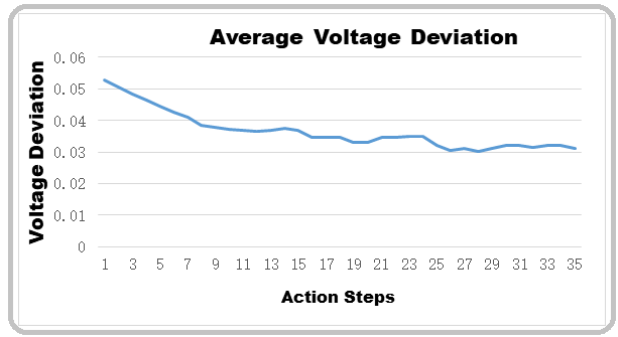

Fig. 4- 3 Variation trend of voltage deviation during single learning

### 4.4.2 Average Voltage Deviation in Multiple Learning

In a single learning, the more times of optimization, the smaller the average voltage deviation value, which indicates that one time learning is effective, but the DQN model can not only reflect at the learning effect of one time; when training many times, there will be certain errors in the learning. The following figure (4-3) shows the average voltage deviation after each learning.

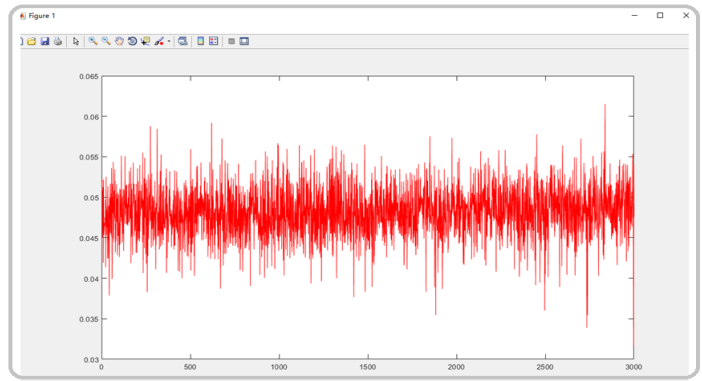

Fig. 4- 4 Average Voltage deviations for multiple learning

Sometimes after learning once, the average voltage deviation will not drop but rise. Each time a section is changed, a total of 3,000 times of section is learned. The values of these average voltage deviations are mostly concentrated between 0.03 and 0.06, with fewer learning failures.

## 4.5 Effect of Nodal Injection in the Distribution Network over Reactive Power Values and Voltage Deviation

Every nodal injection in the network has certain effects on both active power and reactive power that causes imbalance in the network security and distribution network cost pricing in both short term and long term (LRIC). To learn this concept in more practical way, this research has been using the DQN algorithm to train the neural network for the power flow in the network. With the help of IEEE-14 bus node system makes it possible to apply in MATLAB for simulation results.

After learning several power flow sections, test the trained model. The steps are: export the trained model, use the multiple of random number rand (such as 5-fold Rand, -5-fold Rand, etc.) to modify the active and reactive power of the nodes in the IEEE14 node power distribution network model for many times, input the data into the model, check the action strategy given by the model and the effect of implementing the strategy. The average voltage deviation before and after optimization and the comparison line chart of active and reactive power loss of distribution network model are made.

### 4.5.1 Comparison of Voltage deviation before and after Optimization/before and after nodal injection

The post-test model of active and reactive power output of a 5th node was modified using a multiple of the random number rand. In the five tests, the actions selected by DQN model are as follows: the generator voltage at node 3 decreases by 0.01; the generator voltage at node 1 decreased by 0.01; Generator voltage at node 3 decreased by 0.01; the generator voltage at node 1 decreased by 0.01; the generator voltage at node 1 decreased by 0.01. The comparison diagram of average voltage deviation before and after optimization is as follows:

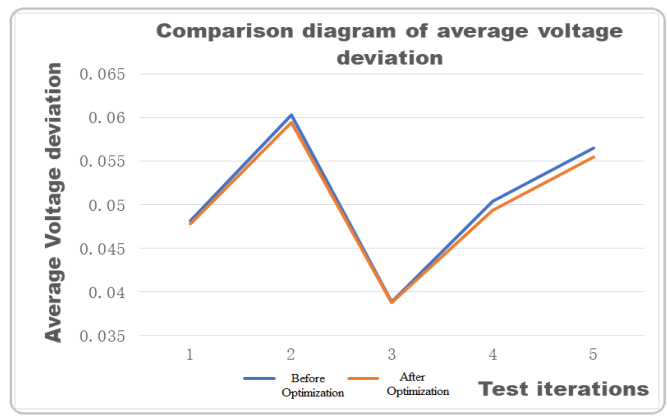

Fig. 4- 5  Change in the voltage deviation before and after the nodal injection

It can be seen that the average voltage deviation after optimization in the figure decreases compared with that before optimization. It shows that DQN model is successful in reducing the average voltage deviation

## 4.5.2 Effect of Nodal Injection in the Distribution Network over Active and Reactive Power Losses.

Network security and network contingencies results in reactive power losses which are applied on this DQN model. The active power and reactive power are compared on the basis of nodal injection i-e results before the nodal injection and after the nodal injection. The figure 4-6 describes the active power losses before and after optimization while on the other hand, figure 4-7 determine the reactive power loss before and after the optimization.

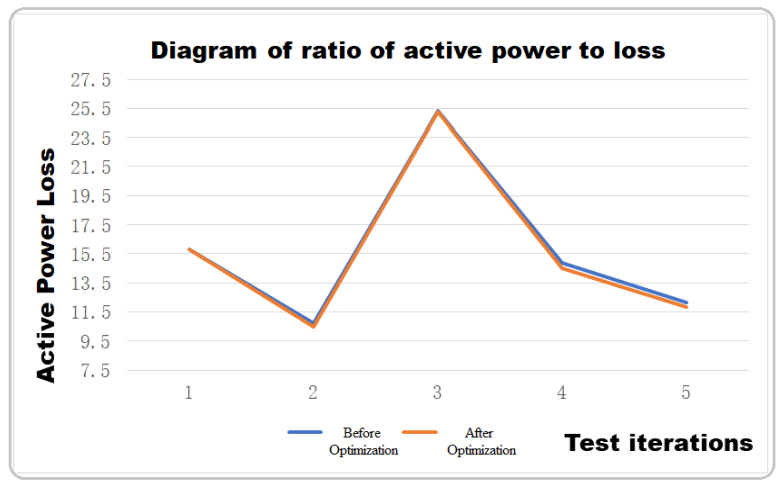

Fig. 4- 6  Compares the active loss before and after optimization

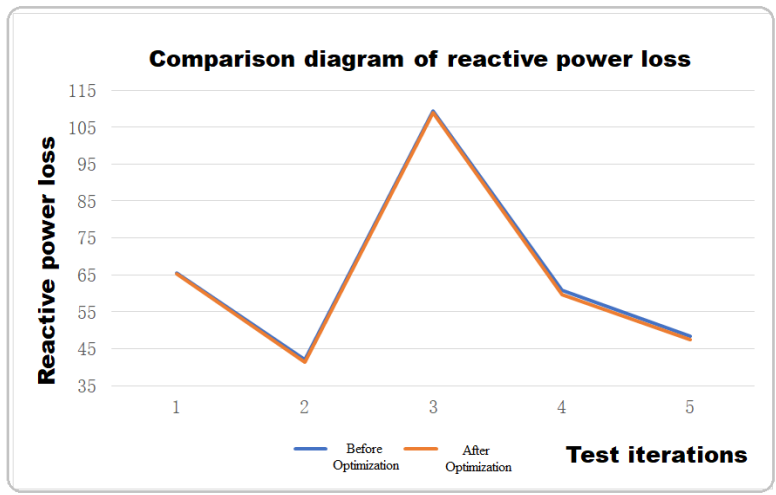

Fig. 4- 7  Compares the reactive loss before and after optimization

Reactive power optimization can not only reduce voltage deviation, but also reduce network loss. The above results show that the strategy selected by DQN model can successfully reduce the active and reactive power loss of distribution network and the complete power grid.

In the results it is shown that the active and reactive power losses before the optimization is different. After optimization, the active and reactive power has different reduced.values.

## 4.6 Optimization of Reactive Power in a Distribution Network

Reactive power optimization is one of the pathway that can reduce the network pricing by regulating the N-1 contingency analysis keeping the network security in a balance situation. This study has proposed this idea and presented methodology for the whole process. It has been already presented in Chapter 3, and most of the successful results have been generated in Chapter 4.

This section of the research highlights the reactive power optimization by learning algorithm (DQN) training the code in MATLAB simulation. The mathematical results are generated with the help of IEEE14 bus node system. The IEEE-14 bus node consists of five generators, four adjustable transformers and one reactive power compensation point. Deep reinforcement learning uses iteration phenomena for reactive power calculation by doing several times iteration following a power flow analysis for the defined values in the program and creates the result.

The iterations are basically training the program using neural networks. The reactive power optimization sheds light on the distribution network pricing and could assume the LRIC pricing in long term manner. To train the program, 14 buses, and 20 branches are used for generating resulting in IEEE-14 system.

The results are generated for at least 1000[th] iterations. The several iterations clearly depict the difference and similarity in reactive power as far as it is optimized in the distribution network. On the basis of those optimized iterative reactive power, the prices of distribution result is calculated. This study analysed different results among them some of the results are shortlisted, analysed with respect to pricing calculation.

### 4.6.1 Iteration 1st of Reactive Power Optimization

The first iteration takes the first repetitive iterative values ranging from one to two. The values and programing code are already defined and presented in chapter 3, section (3.6.3). As discussed the neural network pseudo codes are trained in the MATLAB with the help of using IEEE bus system. The following results shown in table (4-3), (4-4), (4-5), (4-6) (4-7) are calculated for calculating the active power, reactive power values.

Table 4- 3 Data information for the first and second iteration

| Data info | Minimum | Maximum |
|---|---|---|
| Voltage Magnitude | 0.945 p.u. @ bus 2 | 1.070 p.u. @ bus 1 |
| Voltage Angle | -19.54 deg   @ bus 14 | 0.00 deg   @ bus 1 |
| P Losses (I^2*R) | - | 10.65 MW   @ line 1-2 |
| Q Losses (I^2*X) | - | 32.50 MVAr  @ line 1-2 |

Table 4- 4 IEEE 14 Bus Data

| Bus No | Voltage | | Generation | | Load | |
|---|---|---|---|---|---|---|
| | Mag(pu) | Ang(deg) | P (MW) | Q (MVAr) | P (MW) | Q (MVAr) |
| 1 | 1.070 | 0.000 | 277.17 | 201.41 | 2.18 | 4.86 |
| 2 | 0.945 | 4.292 | 40.00 | 237.57 | 22.16 | 16.93 |
| 3 | 1.030 | 15.542 | 0.00 | 133.52 | 97.27 | 21.53 |
| 4 | 0.973 | 12.095 | - | - | 47.85 | -2.51 |
| 5 | 0.973 | 10.304 | - | - | 10.47 | 5.33 |
| 6 | 1.060 | 17.674 | 0.00 | 46.70 | 15.15 | 8.68 |
| 7 | 1.025 | -16.176 | - | - | 1.18 | 4.79 |
| 8 | 1.060 | -16.384 | 0.00 | 24.28 | 2.24 | 3.10 |
| 9 | 1.026 | 18.052 | - | - | 32.35 | 19.60 |
| 10 | 1.023 | 18.316 | - | - | 9.31 | 6.66 |
| 11 | 1.036 | 18.234 | - | - | 5.98 | 2.25 |
| 12 | 1.035 | 18.751 | - | - | 9.31 | 2.88 |
| 13 | 1.029 | 18.631 | - | - | 14.61 | 10.09 |
| 14 | 0.998 | 19.537 | - | - | 19.09 | 9.56 |
| | Total | | 317.17 | 168.35 | 289.13 | 113.76 |

For 14 buses, with different voltages on different buses, the generation capacity values are calculated for individual bus first along with real power/active power (P) and reactive power (Q). The load at both the active power and reactive power calculated values are different; the overall values accumulate towards the total generation capacity values calculated, along with load margin values at the circuit.

Table 4- 5  Branch Data of IEEE 14 bus node system

| Branch Number | From Bus | To Bus | From Bus | Injection | To Bus | Injection | Loss (I^2 * Z) | |
|---|---|---|---|---|---|---|---|---|
| | | | P (MW) | Q (MVAr) | P (MW) | Q (MVAr) | P (MW) | Q (MVAr) |
| 1 | 1 | 2 | 183.78 | 167.62 | -173.13 | -140.50 | 10.646 | 32.50 |
| 2 | 1 | 5 | 91.22 | 28.94 | -86.82 | -15.91 | 4.403 | 18.18 |
| 3 | 2 | 3 | 83.18 | -52.98 | -78.74 | 70.04 | 5.066 | 21.34 |
| 4 | 2 | 4 | 60.83 | -31.76 | -57.83 | 37.74 | 3.003 | 9.11 |
| 5 | 2 | 5 | 46.33 | -29.27 | -44.47 | 31.76 | 1.859 | 5.68 |
| 6 | 3 | 4 | -18.53 | 41.95 | 19.90 | -39.75 | 1.365 | 3.48 |
| 7 | 4 | 5 | -63.84 | 20.37 | 64.47 | -18.37 | 0.633 | 2.00 |
| 8 | 4 | 7 | 34.70 | -12.99 | -34.70 | 15.89 | 0.000 | 2.90 |
| 9 | 4 | 9 | 19.22 | -2.87 | -19.22 | 4.95 | -0.000 | 2.08 |
| 10 | 5 | 6 | 56.35 | -2.81 | -56.35 | 10.16 | -0.000 | 7.35 |
| 11 | 6 | 11 | 9.36 | 8.29 | -9.23 | -8.01 | 0.132 | 0.28 |
| 12 | 6 | 12 | 10.56 | 5.19 | -10.41 | -4.88 | 0.151 | 0.32 |
| 13 | 6 | 13 | 21.28 | 14.37 | -20.90 | -13.61 | 0.388 | 0.76 |
| 14 | 7 | 8 | 2.24 | -20.47 | -2.24 | 21.18 | 0.000 | 0.71 |
| 15 | 7 | 9 | 31.28 | -0.21 | -31.28 | 1.23 | 0.000 | 1.03 |
| 16 | 9 | 10 | 6.10 | 1.01 | -6.09 | -0.98 | 0.012 | 0.03 |
| 17 | 9 | 14 | 12.05 | 4.76 | -11.85 | -4.33 | 0.203 | 0.43 |
| 18 | 10 | 11 | -3.21 | -5.68 | 3.25 | 5.76 | 0.033 | 0.08 |
| 19 | 12 | 13 | 1.10 | 2.00 | -1.09 | -1.99 | 0.011 | 0.01 |
| 20 | 13 | 14 | 7.38 | 5.51 | -7.24 | -5.23 | 0.137 | 0.28 |
| | | | | | | Total | 28.042 | 108.55 |

The nodal injections at each bus and at each 20 branches are given. The power flow in the circuit of IEEE-14 from bus to bus has been presented. From the bus of active power to the injection at reactive powers in each distribution network resulting in different respective values accordingly the data set and the injection measures. The losses are calculated by the formula denoted as the active and reactive power losses in the network. These losses created price fluctuations problem in distribution network for long term and it is one of the challenge LRIC pricing mechanism is facing.

The system summaries for the whole iteration are depicted as below;

Table 4- 6 System IEEE14 Summary

| | |
|---|---|
| Number of Buses | 14 |
| Generators | 5 |
| Committed Gens | 5 |
| Loads | 14 |
| Fixed load | 14 |
| Dispatchable load | 0 |
| Shunts | 1 |
| Branches | 20 |
| Transformers | 3 |
| Inner Ties | 0 |
| Areas | 1 |

Table 4- 7 Detailed IEEE-14 bus System Result Summary

| Number of Categories | P (MW) | Q (MVAr) |
|---|---|---|
| Total Gen capacity | 772.4 | -52.0 to 148.0 |
| Online capacity | 772.4 | -52.0 to 148.0 |
| Generation (actual) | 317.17 | 168.35 |
| Load | 289.13 | 113.76 |
| Fixed load | 289.13 | 113.76 |
| Dispatchable load | -0.0 of -0.0 | -0.00 |
| Shunt (inj) | -0.0 | 17.3 |
| Losses ((I^2 * Z)) | 28.042 | 108.55 |
| Branch Charging | - | 23.8 |
| Total Inter-tie Flow | 0.0 | 0.00 |

The system summary of the whole iteration result evaluated in the MATLAB through IEEE14 bus system by using DQN algorithm technique presents the total number of generation capacity in the network. Actual generation values at active and reactive power generation. Loads are same for the whole process and the shunt values generated because of the resistance and capacitance used in the IEEE14 bus circuit. The losses calculated at active power is 28 and reactive power 108. This is the initial value of reactive power loss in the network for the first iteration.

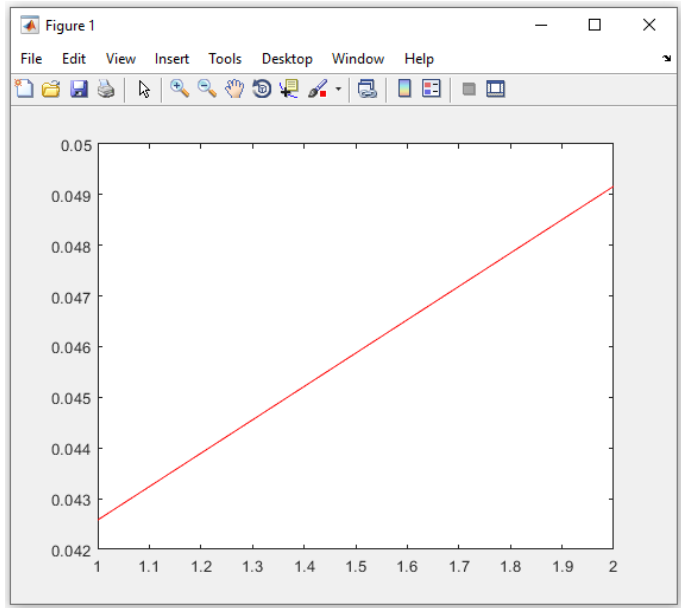

Table 4- 8 Simulation Result of Iteration First to Second via IEEE-14 Bus

The simulation results show a straight line. The line originated form 1 and passing through the minor points and resulted out at 2. The iteration command from MATLAB on IEEE14 bus system is just given once from 1 to 2 periods.

The command generated is to do the learning process in 2 cycles for reactive power optimization. This results points out the reactive power values are still high in the network so as the network pricing still on peak. Further results in the next iteration will optimize reactive power to overcome the pricing difficulty and network security to efficiently reduce reactive power values.

### 4.6.2 Iteration 100th of Reactive Power Optimization

After the first iteration values for reactive power and active power values calculated, this study try to optimize more results for reactive power. The reason for more iteration is to calculate and reduce the reactive power losses in the network.

The more reduced reactive power losses, the more networks will be secure; as a result the network pricing will decrease. This study uses the same methodology of DQN and presented a 100 times optimization of reactive power. The results are calculated and shown in the following tables.

Table 4- 8 Data information for the 100th iteration

| Data info | Minimum | Maximum |
|-----------|---------|---------|
| Voltage Magnitude | 1.013 p.u. @ bus 8 | 1.100 p.u. @ bus 6 |
| Voltage Angle | -19.98 deg  @ bus 14 | 0.00 deg  @ bus 1 |
| P Losses (I^2*R) | - | 6.20 MW    @ line 1-2 |
| Q Losses (I^2*X) | - | 18.92 MVAr @ line 1-2 |

For the mentioned data, the magnitude and angle of voltage in the circuit both at minimum level and maximum level are given respectively different bus nodes. The bus data can be calculated below in table 4-9 from the scenario explained.

Table 4- 9  IEEE 14 Bus Data

| Bus Numbers | Voltage | | Generation | | Load | |
|-------------|---------|---------|---------|---------|---------|---------|
| | Mag(pu) | Ang(deg) | P (MW) | Q (MVAr) | P (MW) | Q (MVAr) |
| 1 | 1.050 | 0.000 | 275.36 | -23.97 | 0.06 | 2.31 |
| 2 | 1.035 | -6.067 | 40.00 | -9.59 | 22.86 | 13.38 |
| 3 | 1.100 | -15.760 | 0.00 | 138.02 | 97.17 | 20.36 |
| 4 | 1.022 | -12.450 | - | - | 48.64 | -1.38 |
| 5 | 1.014 | -10.649 | - | - | 10.63 | 4.31 |
| 6 | 1.050 | -17.706 | 0.00 | 14.10 | 15.24 | 9.71 |
| 7 | 1.056 | -16.659 | - | - | 0.63 | 0.01 |
| 8 | 1.090 | -17.034 | 0.00 | 25.25 | 4.27 | 3.89 |
| 9 | 1.042 | -18.598 | - | - | 33.82 | 18.77 |
| 10 | 1.030 | -18.838 | - | - | 11.74 | 9.12 |
| 11 | 1.031 | -18.522 | - | - | 7.56 | 6.09 |
| 12 | 1.032 | -18.672 | - | - | 6.24 | 2.85 |
| 13 | 1.026 | -18.901 | - | - | 17.44 | 6.99 |
| 14 | 1.013 | -19.978 | - | - | 17.77 | 5.80 |
| | Total | | 315.36 | 143.81 | 294.08 | 102.22 |

The results shown in the table 4-9 calculates the voltage magnitude and the angle; power generation and load in the circuit for each and every single bus used in IEEE14 bus system. The cumulated actual generation values got from the 100[th] iteration results is 315 MW for active power and 143.81 in MVAr for reactive generation. The total circuit load in MW for active power is calculated as 294.08 much higher than the load for reactive power resulted 102.22. These results are at the Bus node for the bus data. The results at the 20 branches are given as below in the table 4-10.

Table 4- 10 Branch Data of IEEE 14 bus node system

| Branch No | From Bus | To Bus | From Bus | Injection | To Bus | Injection | Loss (I^2 * Z) | |
|---|---|---|---|---|---|---|---|---|
| | | | P (MW) | Q (MVAr) | P (MW) | Q (MVAr) | P (MW) | Q (MVAr) |
| 1 | 1 | 2 | 186.23 | -27.00 | -180.03 | 40.19 | 6.198 | 18.92 |
| 2 | 1 | 5 | 89.07 | 0.72 | -85.18 | 10.11 | 3.894 | 16.07 |
| 3 | 2 | 3 | 85.87 | -48.50 | -81.70 | 61.07 | 4.169 | 17.56 |
| 4 | 2 | 4 | 63.54 | -11.37 | -61.30 | 14.57 | 2.239 | 6.79 |
| 5 | 2 | 5 | 47.76 | -3.28 | -46.55 | 3.35 | 1.214 | 3.71 |
| 6 | 3 | 4 | -15.48 | 56.59 | 17.43 | -53.05 | 1.955 | 4.99 |
| 7 | 4 | 5 | -64.64 | 40.11 | 65.38 | -37.78 | 0.740 | 2.33 |
| 8 | 4 | 7 | 38.72 | -3.86 | -38.72 | 6.76 | -0.000 | 2.90 |
| 9 | 4 | 9 | 21.15 | 3.60 | -21.15 | -1.30 | 0.000 | 2.30 |
| 10 | 5 | 6 | 55.71 | 20.00 | -55.71 | -12.55 | -0.000 | 7.46 |
| 11 | 6 | 11 | 10.29 | 5.33 | -10.18 | -5.09 | 0.116 | 0.24 |
| 12 | 6 | 12 | 8.72 | 3.29 | -8.62 | -3.09 | 0.097 | 0.20 |
| 13 | 6 | 13 | 21.47 | 8.32 | -21.15 | -7.69 | 0.318 | 0.63 |
| 14 | 7 | 8 | 4.27 | -20.66 | -4.27 | 21.36 | 0.000 | 0.70 |
| 15 | 7 | 9 | 33.81 | 13.88 | -33.81 | -12.56 | 0.000 | 1.32 |
| 16 | 9 | 10 | 9.19 | 10.28 | -9.13 | -10.14 | 0.056 | 0.15 |
| 17 | 9 | 14 | 11.95 | 5.42 | -11.75 | -4.99 | 0.202 | 0.43 |
| 18 | 10 | 11 | -2.61 | 1.02 | 2.62 | -1.00 | 0.006 | 0.01 |
| 19 | 12 | 13 | 2.38 | 0.24 | -2.37 | -0.23 | 0.012 | 0.01 |
| 20 | 13 | 14 | 6.08 | 0.93 | -6.02 | -0.81 | 0.061 | 0.12 |
| | | | | | | Total | 21.277 | 86.86 |

Finally, the reactive and power results are depicted from this table after a large number of learning algorithm iterating the values for active and reactive power until 100 times.

The mentioned data steps uses the same procedure (table 4-5) of optimizing network reactive power but considering different values, because this optimization lasted for 100 times starting from the starting values compared to the last iteration (1st Iteration).

The power flow is in different direction which is from bus to bus and from bus to the injection nodes. The losses values are calculated with the given formulas and calculate the active power P and reactive power Q values at all the branches and accumulated to the final result. This all process for the whole iterations has been summarizes in the system summary section presented in table (4-11) and (4-12).

The system summaries for the whole iteration are depicted as below;

Table 4- 11 System IEEE14 Summary

| | |
|---|---|
| Number of Buses | 14 |
| Generators | 5 |
| Committed Gens | 5 |
| Loads | 14 |
| Fixed load | 14 |
| Dispatchable load | 0 |
| Shunts | 1 |
| Branches | 20 |
| Transformers | 3 |
| Inner Ties | 0 |
| Areas | 1 |

Table 4- 12 Detail IEEE-14 bus System Result Summary

| Number of Categories | P (MW) | Q (MVAr) |
|---|---|---|
| Total Gen capacity | 772.4 | -52.0 to 148.0 |
| Online capacity | 772.4 | -52.0 to 148.0 |
| Generation (actual) | 315.4 | 143.81 |
| Load | 294.08 | 102.22.6 |
| Fixed load | 294.08 | 102.22 |
| Dispatchable load | -0.0 of -0.0 | -0.00 |
| Shunt (inj) | -0.0 | 27.2 |
| Losses ((I^2 * Z)) | 21.277 | 86.86 |
| Branch Charging | - | 24.1 |
| Total Inter-tie Flow | 0.0 | 0.00 |

The summary of the system is the clear explanation in much simplified shape to differentiate the values from the previous optimization to now. Irrespective of the other values produced in the summary section, the main focus point of this study is analysing, calculating the reactive power values along with the active power. The values of active power P decrease from 26 in first iteration to 21 until 100[th] iteration. Conversely, the reactive power Q values experience huge fall from 107 in first iteration to 86 in the 100[th] learning. This lowering of values highlights the importance of reactive power minimization resulting in lowering the distribution network pricing for long term distribution cost.

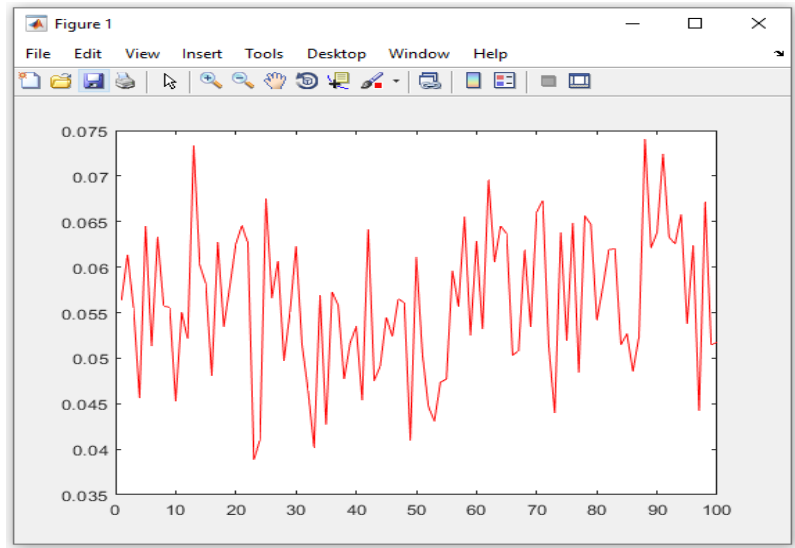

Fig. 4- 8 (a) Simulation Result of Iteration 100th via IEEE-14 Bus

The figure 4-8 (a) is the practical results drawn as an output from the MATLAB simulation that denotes the 100 times optimized values. The graph has several fluctuations due to several times iteration values for distribution network. The simulated graph also highlights the decrease in the losses of reactive power as the fluctuated lines are getting closes to each other. Although the fluctuations has several lower and higher peaks but the closeness in the lines are good notion for distribution network lower pricing.

### 4.6.3 Iteration 1000th of Reactive Power Optimization

In this section, 100th time learning has been performed in the MATLAB for reactive power optimization. The process repeated in 1[st] iteration and 100[th] will remains the same as for this period of optimization technique. This research uses 1000 iteration for optimization of reactive power to reduce the reactive power losses in the network. The losses in the reactive power have been found playing phenomenal role in the network security that causes the instability in the distribution network pricing.

The given values to optimize the 1000 times learning is represented as following in table 4-13.

Table 4- 13 Data information for the 1000th iteration

| Data info | Minimum | Maximum |
|---|---|---|
| Voltage Magnitude | 1.000 p.u. @ bus 8 | 1.080 p.u. @ bus 6 |
| Voltage Angle | -18.32 deg @ bus 14 | 0.00 deg @ bus 1 |
| P Losses (I^2*R) | - | 5.35 MW @ line 1-2 |
| Q Losses (I^2*X) | - | 16.34 MVAr @ line 1-2 |

For the mentioned data, the magnitude and angle of voltage in the circuit both at minimum level and maximum level are given respectively different bus nodes. The bus data can be calculated below in table 4-14 from the scenario explained.

Table 4- 14 IEEE 14 Bus Data

| Bus Numbers | Voltage | | Generation | | Load | |
|---|---|---|---|---|---|---|
| | Mag(pu) | Ang(deg) | P (MW) | Q (MVAr) | P (MW) | Q (MVAr) |
| 1 | 1.070 | 0.000 | 265.46 | 3.72 | 2.69 | 1.68 |
| 2 | 1.045 | -5.442 | 40.00 | 30.32 | 26.57 | 16.93 |
| 3 | 1.030 | -13.800 | 0.00 | 53.78 | 94.93 | 21.81 |
| 4 | 1.012 | -11.245 | - | - | 50.52 | -3.40 |
| 5 | 1.016 | -9.690 | - | - | 8.67 | 5.35 |
| 6 | 1.080 | -16.475 | 0.00 | 37.33 | 11.74 | 8.57 |
| 7 | 1.025 | -16.176 | - | - | 1.35 | 1.23 |
| 8 | 1.000 | -15.006 | 0.00 | -8.98 | 0.72 | 4.99 |
| 9 | 1.038 | -16.747 | - | - | 30.98 | 18.17 |
| 10 | 1.035 | -17.050 | - | - | 10.25 | 7.68 |
| 11 | 1.051 | -17.023 | - | - | 6.86 | 2.70 |
| 12 | 1.055 | -17.459 | - | - | 8.66 | 4.37 |
| 13 | 1.050 | -17.510 | - | - | 16.73 | 8.26 |
| 14 | 1.020 | -18.317 | - | - | 18.13 | 6.30 |
| | | Total | 305.46 | 116.17 | 288.79 | 104.64 |

The IEEE-14 bus system consumes all its 14 buses for the power flow in the network with generators, transformer and capacitor/condenser in the same circuit. The appendix section A-1 derived the practical diagram of this system. As following the sequence of buses in the network structure, the DQN learning method uses deep neural network learning technique to create a base for the power flow in the buses. The power flow in the buses creates active and reactive generation values at their respective nodes along with the load values. The voltages are calculated at the buses nodes including their angles. It can be seen from the table that some of the buses didn't create generation during their power flow because of the higher voltage deviation and negative voltage values. The high fluctuation in the voltages causes the network to not make enough generation. The results are added together for generation and load for both active power and reactive power. Now, the 20 total branches depict the results for nodal injection demonstrated in table 4-15. (Note: The table 4-15 is continued to the next page)

Table 4- 15 Branch Data of IEEE 14 bus node system

| Branch Number | From Bus | To Bus | From Bus | Injection | To Bus | Injection | Loss (I^2 * Z) | |
|---|---|---|---|---|---|---|---|---|
| | | | P (MW) | Q (MVAr) | P (MW) | Q (MVAr) | P (MW) | Q (MVAr) |
| 1 | 1 | 2 | 117.76 | -7.52 | -172.41 | 17.95 | 5.352 | 16.34 |
| 2 | 1 | 5 | 85.02 | 9.55 | -81.54 | -0.53 | 3.483 | 14.38 |
| 3 | 2 | 3 | 77.89 | -7.19 | -75.26 | 13.51 | 2.620 | 11.04 |
| 4 | 2 | 4 | 61.43 | 0.51 | -59.42 | 1.99 | 2.011 | 6.10 |
| 5 | 2 | 5 | 46.53 | 2.11 | -45.39 | -2.31 | 1.137 | 3.47 |
| 6 | 3 | 4 | -19.66 | 18.45 | 20.14 | -18.58 | 0.475 | 1.21 |
| 7 | 4 | 5 | -62.55 | 11.72 | 63.07 | -10.05 | 0.528 | 1.66 |
| 8 | 4 | 7 | 32.62 | 6.09 | -32.62 | -3.94 | 0.000 | 1.80 |
| 9 | 4 | 9 | 18.68 | 2.18 | -18.68 | -0.38 | -0.000 | 1.80 |
| 10 | 5 | 6 | 55.18 | 7.55 | -55.18 | -0.97 | 0.000 | 6.58 |
| 11 | 6 | 11 | 10.48 | 10.52 | -10.30 | -10.14 | 0.180 | 0.38 |
| 12 | 6 | 12 | 10.35 | 5.64 | -10.21 | -5.33 | 0.146 | 0.30 |
| 13 | 6 | 13 | 22.61 | 13.57 | -22.21 | -12.80 | 0.394 | 0.78 |

| 14 | 7 | 8 | 0.72 | 14.31 | -0.72 | -13.97 | 0.000 | 0.34 |
| 15 | 7 | 9 | 30.55 | -11.61 | -30.55 | 12.72 | 0.000 | 1.12 |
| 16 | 9 | 10 | 6.87 | 0.39 | -6.86 | -0.35 | 0.014 | 0.04 |
| 17 | 9 | 14 | 11.38 | 1.39 | -11.23 | -1.06 | 0.155 | 0.33 |
| 18 | 10 | 11 | -3.39 | -7.33 | 3.44 | 7.44 | 0.050 | 0.12 |
| 19 | 12 | 13 | 1.55 | 0.96 | -1.55 | -0.96 | 0.007 | 0.01 |
| 20 | 13 | 14 | 7.03 | 5.49 | -6.91 | -5.24 | 0.123 | 0.25 |
| | | | | | | Total | 16.677 | 68.41 |

These results from the table 4-15 are the final results after long tidy optimization results from the reactive power along with many other factors involved in the network. 1000 times iteration results were chosen as the target iterative point for reactive power in MATLAB programming by DQN learning method. The results surpasses from iteration one to iteration 1000 times. In these results it can be seen that the reactive power and active power values drop to 16 and 68 as compared to the previous iteration results secured from first and hundredth learning periods. The resultant values in the final iteration denote the linear decrease in the value and losses in the reactive power. The decreases in reactive power values are major required results this research has been targeting. In other words, the lower reactive power values mean the lower network security. The lower network security results in lower distribution network pricing. The summary of the iteration are disclosed in the tables (4-15) and (4-16).

The system summaries for the whole iteration are depicted as below;

Table 4- 16 System IEEE14 Summary

| Number of Buses | 14 |
| Generators | 5 |
| Committed Gens | 5 |
| Loads | 14 |
| Fixed load | 14 |
| Dispatchable load | 0 |
| Shunts | 1 |
| Branches | 20 |
| Transformers | 3 |
| Inner Ties | 0 |
| Areas | 1 |

Table 4- 17 Detail IEEE-14 bus System Result Summary

| Number of Categories | P (MW) | Q (MVAr) |
|---|---|---|
| Total Gen capacity | 772.4 | -52.0 to 148.0 |
| Online capacity | 772.4 | -52.0 to 148.0 |
| Generation (actual) | 305.4 | 116.1 |
| Load | 288.8 | 104.6 |
| Fixed load | 288.8 | 104.6 |
| Dispatchable load | -0.0 of -0.0 | -0.00 |
| Shunt (inj) | -0.0 | 32.2 |
| Losses ((I^2 * Z)) | 16.60 | 68.15 |
| Branch Charging | - | 24.5 |
| Total Inter-tie Flow | 0.0 | 0.00 |

The total generation capacity has been kept fixed in the programming code defined for IEEE14 nodes. For the mentioned results, the generation capacity is found comprising both active and reactive power. The shunt resistance values has changed and increasing as compared to the first iteration to the final iteration. The branch charging voltage has been found decreasing which is a good sign for the reactive power lowering values. The active power P and reactive power Q values drops down compared to the last optimized learned data. The following figure 4-9 show the results for the summarized data.

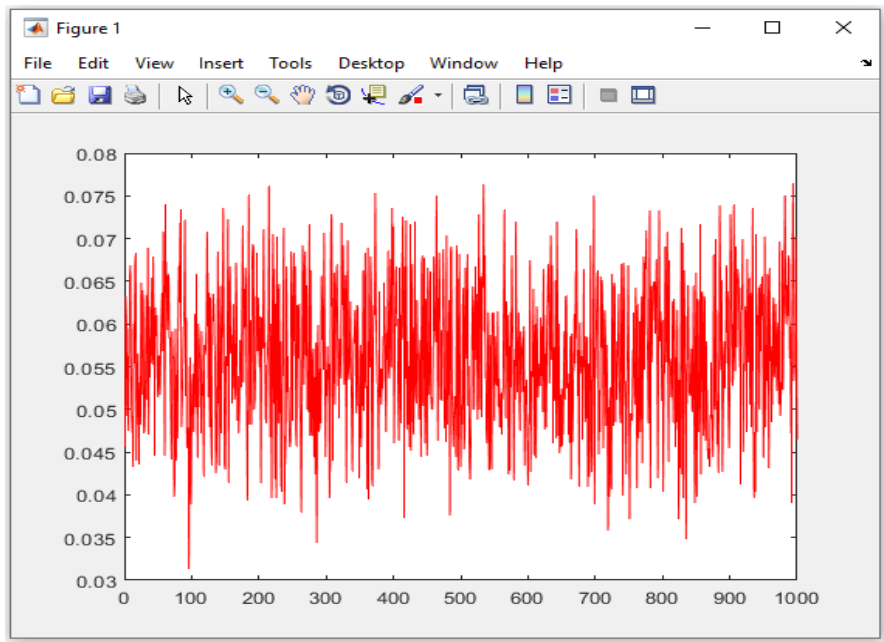

Fig. 4- 9 Simulation Result of Iteration 1000th via IEEE-14 Bus

The simulated results after the final iteration can be reflected from the figure 4-8. According to the resulted figure, maximum of 1000 times reactive power has been simulated considering active power, voltage voltages, effect of nodal injection at the node branches. The fluctuated graph results determine the closeness between the zigzag lines that denotes the maximum decrease in the reactive power loss. The more the fluctuated lines nearer to each other, the more better will be the reactive power values with lower losses. The lower losses in the reactive power, more the network will be considered secure with highest network security factor. As discussed, the larger network security or security factor can reduce the LRIC pricing.

The results are repeated several times until the required proposed data are achieved. The iteration never stops during optimizing reactive power from the starting point of first iteration to the suggested 1000 iteration of repetitive process. During this duration, the repetitive period takes longer period of time to optimize the reactive power and get back to the target point set in the programming learning code in MATLAB. The periodic process passes through every node and branch of the network to accurately configure and optimize the reactive power values. The longer process may be tidy but it is much helpful to remove the network security problem.

The three iteration results are the shortlisted results to practically propose and prove the method of reactive power optimization through the DQN learning neural network using IEEE14 as the best possible approach for reducing reactive power values. The lowest reactive power is an indirect definition of lower losses in the network. The lower network losses mirror lower distribution network pricing which is a future challenge for LRIC pricing.

This study approaches and tries to solve and propose the lower network cost challenge faced by long run incremental cost (LRIC). In other words, this research provides solution for LRIC pricing to overcome the network security for current and future customer in Great Britain. Further optimization results are shown the appendix section. Appendix C determines more optimization results accordingly the process applied on first, hundredth and one thousands iteration.

## 4.7 LRIC Charges for Active Power P and Reactive Power Q comparing network security

After the reactive power optimization process, this study sheds light in the change of electricity prices and impact on security factor after the application of optimization methodology. The results with respects to IEEE14 bus system are shown in the following figures.

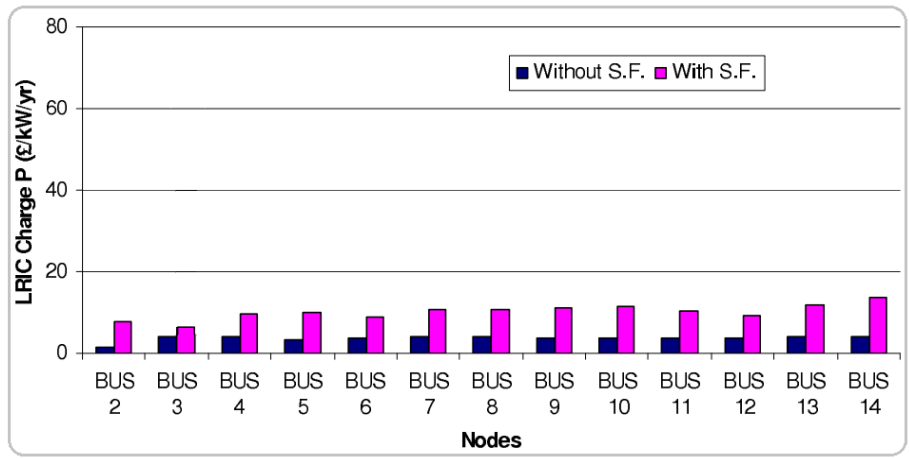

Fig. 4- 10 LRIC pricing comparison for active power P with and without network security factor analysis

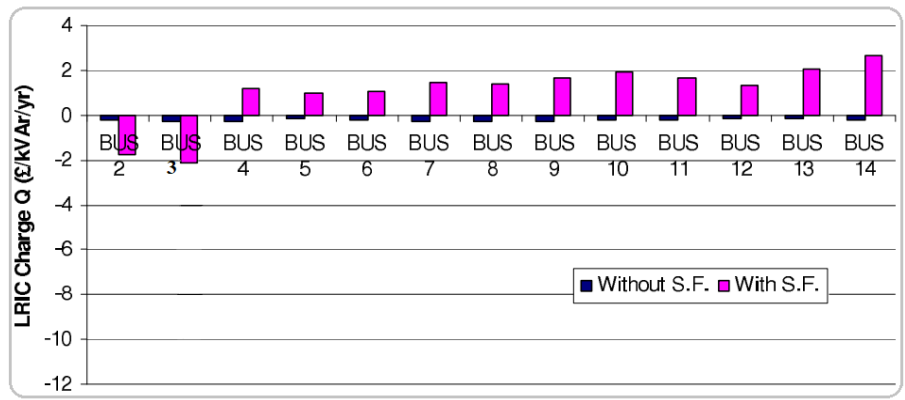

Fig. 4- 11 LRIC pricing comparison for reactive power Q with and without network security factor analysis

Figure 4-9 shows the LRCIC charges derived for active power through IEEE14 system and DQN algorithm. These charges compared the result of active power after many times optimization as a result decrease in the network security. It can be seen that charges without security factor is pretty high as compared to the pricing with considering the network security factor. Alternatively, figure 4-10 demonstrates the long run incremental charges for reactive power with and without network security. Without considering network security factor (S.F), the whole charges values are minor negative as can be seen in the figure 4-10. These negative values suggests the presence of huge number of reactive power in the network system that causes the network to not withstand the full N-1 contingent situation. At bus 2 and 3, the security factor values are high negative price because of the reactive power cross-flow originated in other buses like 1, 3 and 5. The same is true for bus 3 peak negative price due to the reason of reactive power reverse flow originated in buses 1, 4, 6 and 7.

# Chapter 5 Reforming Incremental Distribution Network in China: Lesson learned from UK

This chapter of the study highlights the current china's incremental distribution network. This study makes a base and provides suggestion for China's incremental distribution network from the experience of the UK distribution network pricing methodologies. This research also discusses the reforms structure in the power system and distribution network system along with the network pricing calculation methods.

## 5.1 Comparison overview of current China's Incremental distribution network and UK

China's power distribution network is in reforming stage. China is putting efforts huge and attention towards bringing innovation and reforms in regulating the incremental distribution network. China being the world largest producer of renewable energy resources made it complementary to adjust the embedded generation in the distribution network. The UK electricity market system as discussed in the previous study is the most advanced and mature; the above research make a visionary platform for Chinese power network system guided by the UK incremental distribution network to efficiently promote the clean energy generation and accurately regulate the incremental distribution network security and its pricing. Albeit, playing the leading role in the world for the clean energy production, it is experiencing major challenges dealing with the distribution network and the distribution network system.

The Great Britain has a decentralized market for all type of power system cost especially for the distribution network. The UK network system consists of 14 DNOs which are regulated by OFGEM. In China, the main China State grid and China Southern power grid are the responsible authorities for the distribution network. To get rid of the monopoly system, China's power system has launched more than 350 Pilot projects in last three to four years for efficient distribution network. These Pilot projects are generating plant as well as distribution network by making chain of small Microgrids in the distribution network system to cope with the incremental distribution pricing. The mentioned Pilot projects are built in all over China and according to the China's National Energy administration (NEA), the number of Pilot projects will exceed 700 until the end of 2020.

## 5.2 Guidelines for Reform in China's Incremental distribution network pricing

To formulate local power grids and incremental distribution networks in a scientific and reasonable manner (hereinafter referred to as "distribution networks"); Distribution price, promote the healthy development of distribution network business, according to the electricity pricing Law of the People's Republic of China. To practically design the law, the CPC central committee and state council urged the electricity sector to promote efficient price mechanism. Several opinions on further deepening the reform of the power system were established in 2015 (documents released by the Chinese government). Therefore, National energy administration (NEA) and national development and reform commission (NDRC) are taking it serious to standardize the reform of the incremental power distribution.

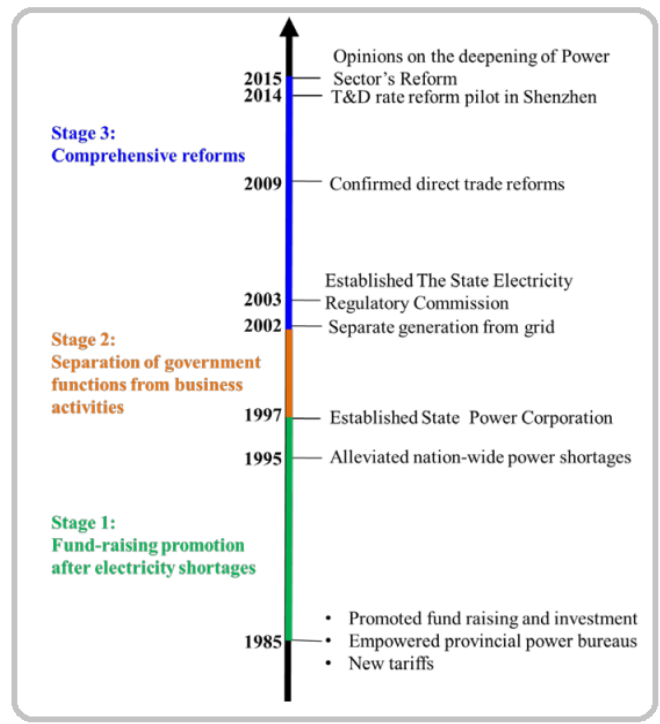

Fig. 5- 1 The timeline for reform progress in China's Power system

### 5.2.1 General Requirements

In accordance with the overall requirements of deepening the reform of the power system the basic idea of local governments or other entities to build and operate local power grids. The three main requirements are discussed as;

1.  First, the establishment of mechanisms and reasonable pricing: Through system and rule construction, the both mentioned terms to provide high scientific government pricing, strengthen the cost and price supervision of distribution network; and specification with the price behavior of power grid enterprises to forms a scientific, reasonable, open and transparent distribution system with effective incentives price, promote fair competition in the electricity market.

2.  The second is to make up for the combination of cost and constrained incentive: On the basis of strict cost supervision and examination, to determine the distribution price in accordance with the principle of making up the reasonable costs of distribution network enterprises and obtaining reasonable benefits; Promote the healthy and sustainable development of distribution network and provide safe and reliable power services. Excitation constraint mechanism promotes distribution network enterprises to improve efficiency, reduce costs as low as possible for users to provide quality power distribution services.

3.  Third, to combine fairness, openness and equal burdens: Distribution network and provincial grid have equality the dominant position of the market. Provincial power grids should be open to local power grids and incremental distribution networks without discrimination. The distribution network shall be open to the power selling company without discrimination. Distribution network enterprises shall follow the same principle and standard undertake policy to cross subsidy.

## 5.3 Allocation of Distribution Network Pricing in China

The price of electricity used by power users within the distribution network shall be determined by the on-grid price or the market price of electricity. The sum of the distribution price of distribution networks undertaken by users and the transmission and distribution price of the power grids at the next higher level shall not be high. It is directly connected to the current provincial power transmission and distribution price corresponding to the same voltage grade. The provincial price department in charge shall, in accordance with the situation of the province, fully solicit relevant enterprises. Consideration is given to the regional electricity price, provincial power grid transmission and distribution price, wholesale price, sales price, etc.

For the distribution network projects whose investment subjects are determined through bidding, the bidding method shall be adopted. Price method determines the distribution price. Bidders shall make investment scale, power distribution capacity and supply at the same time. Electrical reliability, service quality, line loss rate and other commitments are some of the supply standard for the distribution price. If the prescribed power supply service standard is not up to the agreed standard, it shall be subject to supervision and assessment

For distribution network projects in which the investment subjects is determined by non-tendering, choices might be made for one or more of the three pricing methods. The permitted income method, the maximum price method and the scale competition method.

## 5.3.1 Permitted Income Method

Provincial price authorities shall determine power distribution in the competent department of energy. Network planning investment and project owners determine the investment plan, with reference to the provincial power grid transmission and distribution price. The permitted costs, permitted profits and taxes within the price shall be determined within the period of supervision.

## 5.3.2 The Maximum Price Method

First of all, it is calculated according to the method of "permitted cost plus reasonable benefit". The distribution price of a certain distribution network shall refer to the distribution price of other distribution networks with comparability to determine the distribution network by combining performance appraisal indexes such as power supply reliability and service quality. The maximum price of distribution shall be determined by the distribution network enterprise and the provincial price shall be reported. For the record of the competent authority, the authority will encourage local governments to explore ways to set maximum prices in line with the consumer price index.

## 5.3.3 The Scale Competition Method

This method calculate the pricing according to the method of "Permitted cost plus reasonable profit". This means that it is based on the distribution price of a certain distribution network, and then according to the calculation of the distribution price of the distribution network including different provinces. The weighted average distribution price of the distribution network is used to determine the distribution price of the distribution network. At first, the distribution network can be given a higher weight in the supervision period, where the distribution network difference is small. The average advanced level of the same type of distribution network society can also be taken as the benchmark and benchmarking can be made according to the classification of provinces distribution price.

## 5.4 Adjustment Mechanism of Distribution Network Pricing in China

Distribution network distribution price adjustment should be clear about the price supervision cycle, and in the transition phase of the price. Grid connection, and provincial power grid transmission and distribution price pricing method (trial) to establish smooth Processing mechanism, regular check mechanism and assessment mechanism. By doing so;

Firstly, it is necessary to define the supervision period of distribution price. Normally, the government sets the regulatory cycle for distribution prices which is three years in principle. The term of validity of power distribution price shall be determined by tendering, and shall be stipulated in the power distribution project contract.

Secondly, the proper linkage of prices at the transition stage. The electricity price of transmission and distribution settled by the distribution network is temporarily based on the current provincial grid corresponding to its access voltage level. When the existing small hydropower generation projects in the square grid area conduct close transactions with power users, the user shall only pay the distribution price of the voltage level used and shall not undertake the transmission of the previous voltage level.

Third, making good connection with distribution prices of existing local power grids. In accordance with the principle of respecting history and making proper connections, and on the premise of not increasing cross-subsidies, set the distribution price of local power grid, and compare it with the current provincial power grid transmission and distribution price, wholesale price, etc.

Fourth, to encourage the establishment of incentive mechanisms. This is one of the important lesson learned from the distribution network pricing method of UK. This concept will encourage enterprises to improve operational efficiency and reduce distribution costs.

Fifth, the settlement system. The settlement price between the distribution network and the provincial grid shall be based on the corresponding voltage of the current provincial grid. Distribution network enterprises can choose the classification node independently according to the actual situation and Calculate electricity price or comprehensive settlement price and settle electricity price with provincial power grid enterprises at different voltage levels. If the difference between the distribution price of electricity and the actual cost is too high, the provincial price administrative department may, according to the actual situation, set the price at the same time.

# Chapter 6 Conclusion

Electricity distribution network is considered as one of the important fundamental part of the power system. In modern digitized power system, it has gained much attention due to the injection of wide variety of distributed generation into the distributed generation. Due to the increase in the global warming, energy experts around the world warned the grid authorities and energy producers to encourage them producing more greener energy. In this context, china is leading the role of producing more clean energy compared to other countries. UK is one of the country focused on shifting their demand and supply to renewable energy generations resources.

This study pointed out the advantages along with the challenges created by the continuous injection of DGs sources into the distribution network. The advantages in the sense that; the embedded generation or nodal injection of renewable energy resources in the network is very helpful to match the supply and demand of the consumers because the traditional energy generation from coal and other resources is considered not enough due to highly use of energy resources in the modern era. The advancement in power system by inclusion of new resources into the distribution grid or smart grid helps to build a decentralize electricity market. The new clean DGs are also very helpful for the clean environment and zero carbon. On the other hand, it creates lot of challenges for the distribution network operators. Network security is one of the major challenge faced by the network due to the reason because it directly affects the pricing of the distribution network. Reactive power is a type of factor that most of the distribution network didn't highly regard its impact over the distribution network pricing in the long term incremental cost (LRIC).

This research proposes reactive optimization methodology to regulate and reduce the distribution network pricing in long term and to create balance in the distribution network system caused by the network security due to the new nodal injection of DGs. For this purpose, it presented Deep reinforcement learning algorithm (DQN) which uses neural network to reinforce the learning code. An IEEE14 bus system is modeled to evaluate the power flow in the network. The network security creates contingency in the network and made it N-1, N-2,……Nn. The more network is contingent the more security factor will be higher which will results in the higher pricing. This study concludes that network security is an essential element of a distribution network. If the network security of a network is stable, it will build a stable network. The stable distribution network can make it easier for the distribution network companies to distribute cheap electricity cost among the customers.

China's network distribution system is in evolution stage and many reforms are discussed to tackle the incremental distribution network. This study made some suggestions from the UK experience of handling distribution network system. The lesson learned from UK in this research by China is to produce, highlight and promote major reforms in the distribution network system and long run incremental distribution costs. This research concludes that the efforts made by china so far to tackle the distribution network challenge is highly satisfactory. This study learned more than 300 Pilot Projects deployed in China that will help to add energy generated from renewable sources into the main grid. This will creates competition for the grid companies by challenging monopoly system and efficient distribution of electricity with accurate cost. The thesis concludes that if large numbers of the Pilot projects are deployed around the country as faster it can; the easier will be to handle the challenge of network security. The network security will make the network zero contingent and trustworthy as well as highly cost effective.

## Future Perspectives of the Research

China's incremental distribution network structure is still in the development, transformation and reform stage. The network security will stay an alarming situation for some span of time but the increase number of Pilot Projects around the country and strengthening reforms could reduce the probability of network security in near future. This research provides a platform recognising the network contingency and solves the network security problem for charging consumers on the basis of their true network use. The presented model of reactive power optimization through DQN algorithm is a model concept but it still has some discrepancies. To optimize reactive power and voltage deviation calculation in the distribution network is a very tricky technique, and solving through DQN algorithm could be little slower. The DQN algorithm requires lot of time to optimize every single reactive power values by several times iteration which is time consuming. The huge number of renewable energy addition in the distribution network need to deal in a right specific time to meet the consumers demand without time delay. Therefore, this method needs further improvement to optimize the reactive power in a timely manner which is the major challenge for this research problem in the future. In addition, to deal with the Pilot Projects distribution pricing in China on provincial based is another key challenge for this research work since the increasing number of Pilot Projects around the whole China.

# APPENDIX

## Appendix A

The proposed IEEE 14-Bus system is as given as;

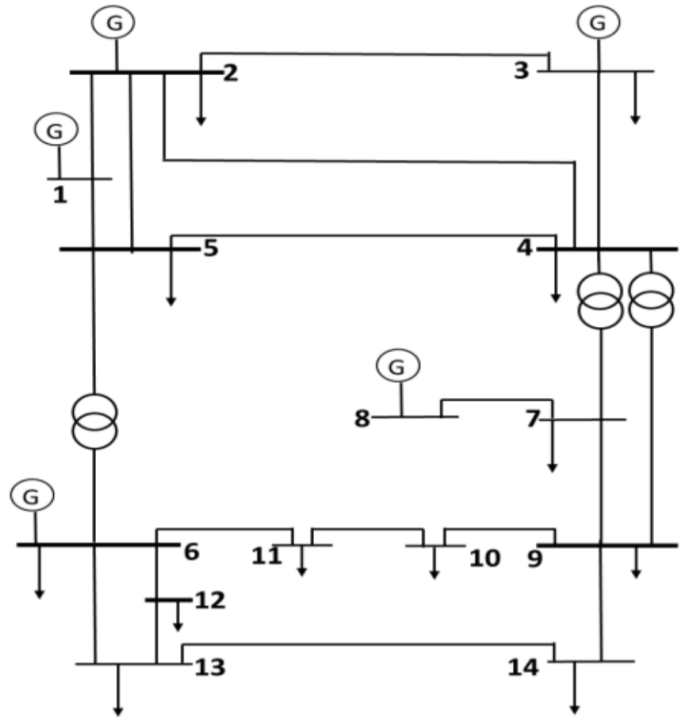

Fig. A- 1 The IEEE 14 bus system

Table A- 1 Maximum allowed loading level (MALL) at specified branches

| Branches | From Bus | To Bus | Base Loading level MVA | Maximum Allowed Loading Level (MALL) MVA | Security Factor (S.F) |
|---|---|---|---|---|---|
| 1 | 4 | 7 | 89.67 | 67.93 | 1.32 |
| 2 | 4 | 9 | 60.06 | 30.80 | 1.95 |
| 3 | 5 | 6 | 100.20 | 73.68 | 1.36 |

The table depicts that MALL has close relation with the rated capacity as both are growing in ascending order but for all the branches at the IEEE 14- bus system, the security factor remains in the range of 1.00 to 7.00. It is further sketched in the figure A-2 of the appendix section.

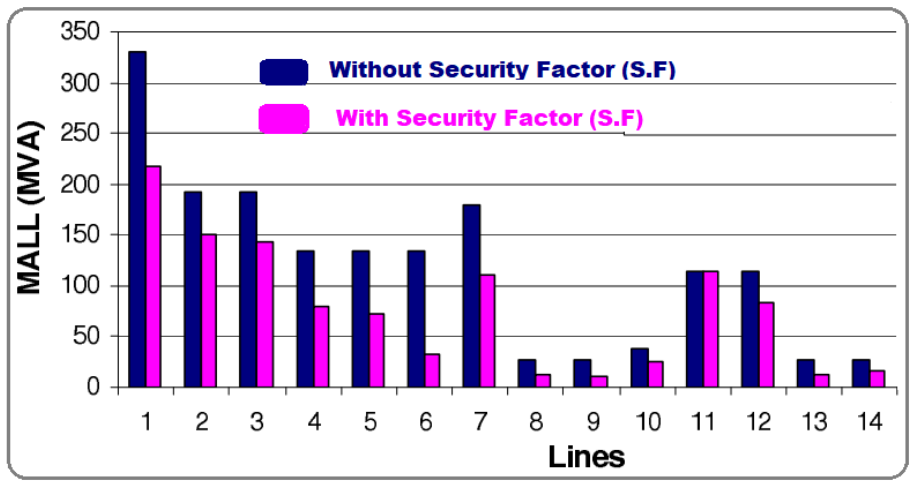

Fig. A- 2 MALL values considering without and with security factor

Without security factor, the network is more N-1 contingent at the buses nodes with lagging maximum allowed loading level.

# Appendix B

In this appendix B, the picture diagram of Neural Network window is shown below;

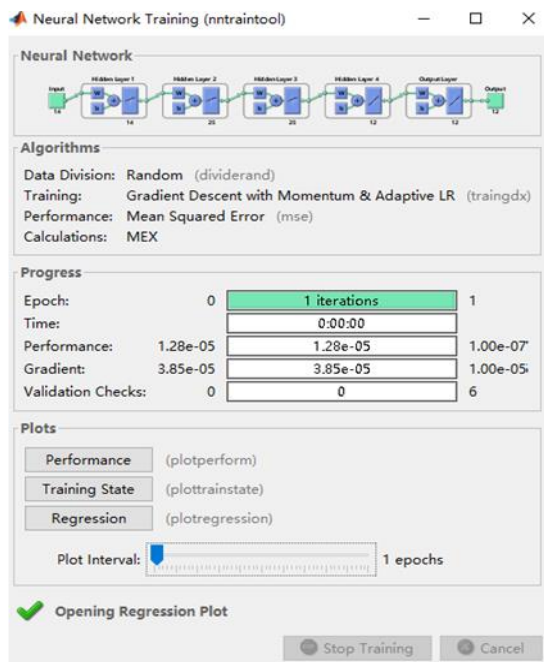

Fig. B- 1 The Neural Network Training Window

After 1000 to 2000 times of training with one of the power flow sections, the NN train tool window in the training of the neural network in the program is opened, from which the neural network model can be read, and the parameters of the training model can be observed through images. Neural Network is a model diagram of the neural network, with input at the front and four hidden layers in the middle. The hidden layer is followed by an output layer and output layer. The number below the middle layers represents the number of neurons. Next is the arithmetic is data division, Random denotes the use of random numbers to divide the target into three groups which are train, validation, test. Training denotes that the neural network for learning training is gradient descent and adaptive, Mean Squared Error denotes that performance is represented by mean square error.

# Appendix C

Reactive Power Iteration values for 500<sup>th</sup> time periods is shown here below

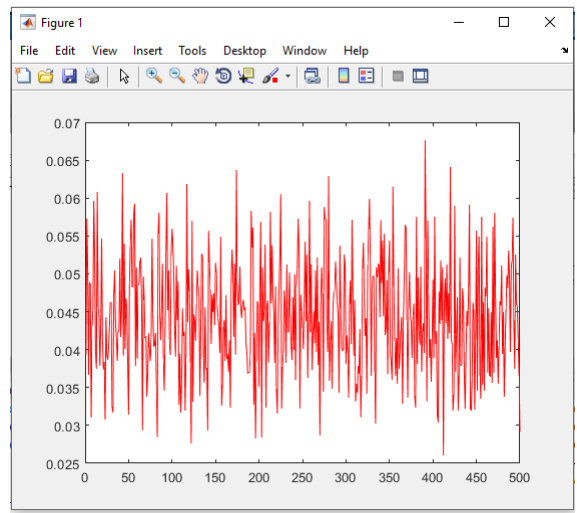

Appendix C- 1 Simulation Result of Iteration 500th via IEEE-14 Bus

Reactive Power Iteration values for 800<sup>th</sup> time periods is shown here below

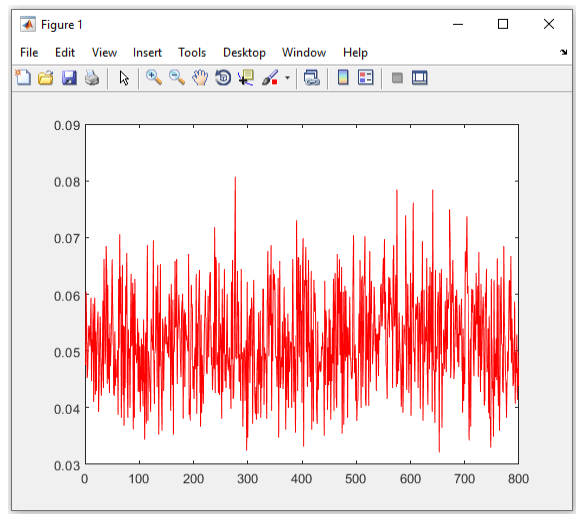

Appendix C- 2 Simulation Result of Iteration 800th via IEEE-14 Bus

# Papers and other achievements published during Master's studies

## Published academic papers

# **Acknowledgement 致谢**

Firstly, I would like to thank my Supervisor, Prof. Wang Peng for his valuable guidance and the time he took to offer me encouragement, insight, and direction for my research work. His expertise in the field of electricity market and distribution network pricing helped me a lot in the completion of my Master's thesis. He has been my mentor, guide and the driving force behind this whole research work. His positive attitude and friendly behavior will always be an inspiration for me.

I would also like to extend my gratitude to all my lab mates' friends Xiye Liye, Zhang Pengyu, Sunhua Kai, Hong Xiaofeng, Cao Yujiye, Li Meng and Jiang Kai for their unconditional support and help during my whole research period. Their willingness to coordinate with me made my research much easier. In addition, they always treated me like their family member due to the reason I never felt to be staying away from home. I wish all of them a successful career and happy life.

In the end, I am greatly thankful to my parents for their continuous support and well wishes during my academic duration. Thanks to North China Electric Power University, Beijing for providing me a splendid platform to undertake postgraduate studies in a timely manner.